\documentstyle[prl,epsf,aps]{revtex}
  
\newtheorem{theorem}{Theorem}
\newtheorem{remark}{Remark}

\newtheorem{lemma}[theorem]{Lemma}
\newtheorem{proposition}{Proposition}
\newtheorem{corollary}{Corollary}
\newtheorem{conjecture}{Conjecture}
\newtheorem{assumption}{Assumption}

\newcommand{\bpr} {\noindent {\bf Proof }}
\newcommand{\epr}{$\Box$}
\newcommand{\bd}{
\begin{document}}
\newcommand{\ed}{\end{document}}
\newcommand{\beq}{\begin{equation}}
\newcommand{\eeq}{\end{equation}}
\newcommand{\bef}{\begin{figure}}
\newcommand{\enf}{\end{figure}}
\newcommand{\bea}{\begin{eqnarray}}
\newcommand{\eea}{\end{eqnarray}}
\newcommand{\baR}{\begin{array}}
\newcommand{\eaR}{\end{array}}
\newcommand{\bth}{\begin{theorem}}
\newcommand{\eth}{\end{theorem}}
\newcommand{\bhyp}{\begin{assumption}}
\newcommand{\ehyp}{\end{assumption}}
\newcommand{\bp}{\begin{proposition}}
\newcommand{\ep}{\end{proposition}}
\newcommand{\bco}{\begin{corollary}}
\newcommand{\eco}{\end{corollary}}
\newcommand{\bconj}{\begin{conjecture}}
\newcommand{\econj}{\end{conjecture}}
\newcommand{\ble}{\begin{lemma}}
\newcommand{\ele}{\end{lemma}}
\newcommand{\bR}{\begin{remark}}
\newcommand{\eR}{\end{remark}}
\newcommand{\bc}{\begin{center}}
\newcommand{\ec}{\end{center}}
\newcommand{\ben}{\begin{enumerate}}
\newcommand{\een}{\end{enumerate}}
\newcommand{\bit}{\begin{itemize}}
\newcommand{\eit}{\end{itemize}}
\newcommand{\su}{\section}
\newcommand{\ssu}{\subsection}
\newcommand{\sssu}{\subsubsection}
\newcommand{\nid}{\noindent}
\newcommand{\nnb}{\nonumber}

\newcommand\bbbr{{\sf I\!R}}
\newcommand\bbbc{{\sf I\!\!C}}
\newcommand\bbbn{{\sf I\!N}}
\newcommand\bbbh{{\sf I\!H}}
\newcommand\bbbz{{\sf Z\!\!Z}}                                            
\newcommand\bbP{{\sf I\!\!P}}
\newcommand\cA{{\cal A}}
\newcommand\cB{{\cal B}}
\newcommand\cC{{\cal C}}
\newcommand\cD{{\cal D}}
\newcommand\cE{{\cal E}}
\newcommand\cF{{\cal F}}
\newcommand\cG{{\cal G}}
\newcommand\cH{{\cal H}}
\newcommand\cI{{\cal I}}
\newcommand\cJ{{\cal J}}
\newcommand\cK{{\cal K}}
\newcommand\cL{{\cal L}}
\newcommand\cM{{\cal M}}
\newcommand\bcM{{\bar{\cal M}}}
\newcommand\cN{{\cal N}}
\newcommand\cO{{\cal O}}
\newcommand\cP{{\cal P}}
\newcommand\tcP{\tilde{\cal P}}
\newcommand\cQ{{\cal Q}}
\newcommand\cR{{\cal R}}
\newcommand\cS{{\cal S}}
\newcommand\cW{{\cal W}}
\newcommand\cU{{\cal U}}
\newcommand\cV{{\cal V}}
\newcommand\cT{{\cal T}}
\newcommand\cX{{\cal X}}
\newcommand\cZ{{\cal Z}}
\newcommand\cl{{\cal l}}
\newcommand\cn{{\cal n}}
\newcommand\ci{{\iota}}

\newcommand\hX{\hat{\bX}}
\newcommand\hY{\hat{\bY}}
\newcommand\hF{\hat{\bF}}
\newcommand\hm{\hat{\mu}}
\newcommand\bE{{\bf E}}
\newcommand\bF{{\bf F}}
\newcommand\bJ{{\bar J}}
\newcommand\bX{{\bf X}}
\newcommand\bY{{\bf Y}}
\newcommand\bU{{\bf U}}
\newcommand\bZ{{\bf Z}}
\newcommand\hZ{{\hat \bZ}}
\newcommand\bb{{\bf b}}
\newcommand\be{{\bf e}}
\newcommand\bk{{\bf k}}
\newcommand\bh{{\bf h}}
\newcommand\bhl{{\bar \phi}_L}
\newcommand\Bhl{{\bar \Phi_L}}
\newcommand\bi{{\bf i}}
\newcommand\bj{{\bf j}}
\newcommand\bl{{\bf l}}
\newcommand\bn{{\bf n}}
\newcommand\bq{{\bf q}}
\newcommand\br{{\bf r}}
\newcommand\bS{{\bf S}}
\newcommand\bT{{\bf T}}
\newcommand\bbq{{\bar{\bq}}}
\newcommand\bu{{\bf u}}
\newcommand\bbUL{{\bar{\bU}_L}}
\newcommand\bv{{\bf v}}
\newcommand\bx{{\bf x}}
\newcommand\bXLp{{\bar{X}_L^+}}
\newcommand\bXLpi{{\bar{X}_{L,i}^+}}
\newcommand\bbXLp{{\bar{\bX}_L^+}}
\newcommand\by{{\bf y}}
\newcommand\bz{{\bf 0}}
\newcommand\bw{{\bf w}}
\newcommand\btX{{\bf \tilde{X}}}
\newcommand\bxi{{\bar{\xi}}}
\newcommand\sxi{{\sigma_\xi}}

\newcommand\ta{{\tilde a}}
\newcommand\tb{{\tilde b}}
\newcommand\ti{\tilde{\i}}
\newcommand\tx{\tilde{x}}
\newcommand\txT{\tilde{x}_T}
\newcommand\txTk{\tilde{x}_{T,k}}
\newcommand\tbx{\tilde{\bx}}
\newcommand\tbxT{\tilde{\bx}_T}
\newcommand\ctbxT{\left[\tbx\right]_T}
\newcommand\tu{\tilde{u}}
\newcommand\tuT{\tilde{u}_T}
\newcommand\tbu{\tilde{\bu}}
\newcommand\tbuT{\tilde{\bu}_T}
\newcommand\ctbuT{\left[\tbu\right]_T}
\newcommand\ctaT{\left[\ta\right]_T}
\newcommand\tv{\tilde{v}}
\newcommand\tvT{\tilde{v}_T}
\newcommand\tbv{\tilde{\bv}}
\newcommand\tbvT{\tilde{\bv}_T}
\newcommand\ENT{E^{(N)}_{\ctbuT}\left[\right]}
\newcommand\txi{\tilde{\xi}}
\newcommand\tmu{\tilde{\mu}}
\newcommand\tmuT{\tilde{\mu}_T}
\newcommand\tnu{\tilde{\nu}}
\newcommand\nul{\nu_L}
\newcommand\mul{\mu_L}
\newcommand\mulo{\mu^{(0)}_L}
\newcommand\mult{\mu^{(t)}_L}
\newcommand\tbeta{\tilde{\beta}}
\newcommand\tom{\tilde{\omega}}
\newcommand\bom{\bar{\omega}}
\newcommand\bt{\bar{\theta}}
\newcommand\st{\sigma_\theta}
\newcommand\sdt{\sigma^2_\theta}
\newcommand\tml{\tilde{\mu}_L}
\newcommand\ml{m_L}
\newcommand\hml{\hat{\mu}_L}
\newcommand\hmlo{\hml^{(0)}}
\newcommand\hmlt{\hml^{(t)}}
\newcommand\bel{\bar{e}_L}
\newcommand\bxl{\bar{X}_L}
\newcommand\bpxl{\bar{X}^+_L}
\newcommand\brl{\bar{r}_L}
\newcommand\boml{\bar{\omega}_L}
\newcommand\SLp{\Sigma_\Lambda^+}

\newcommand\bcm{\bar{\cM}} \newcommand\cm{\cM}

\newcommand{\deq}{\stackrel {\rm def}{=}}
\newcommand{\peq}{\stackrel {\rm p.s.}{=}}
\newcommand\Ppto{\stackrel{\bbP ps}{\to}}
\newcommand\tow{\stackrel{faiblement}{\to}}

\newcommand{\sep}{\; \,}
\newcommand{\D}{\displaystyle}
\newcommand{\T}{\textstyle}
\newcommand{\etc}{etc $\dots$}
\newcommand{\etal}{etc $\dots$}
\newcommand\dLambda{\partial \Lambda}
\newcommand\dL{^\partial \Lambda}

\def\Appendix{\section*{APPENDIX}}

\baselineskip18pt

\bd

\baselineskip18pt

\title {Self-Organized
Criticality and Thermodynamic formalism.}

\author {B. Cessac
\thanks{Institut Non Lin\'eaire de Nice, 1361 Route des
Lucioles, 06560 Valbonne, France}
Ph. Blanchard
\thanks{University of Bielefeld, BiBoS, Postfach 100131, D-33501,
Bielefeld, Germany}
T. Kr\"uger
\thanks{University of Bielefeld, BiBoS, Postfach 100131, D-33501,
Bielefeld, Germany and Technische Universitaet, Str. des 17 July 135, 1062
3, Berlin, Germany}
J.L. Meunier,\thanks{Institut Non Lin\'eaire de Nice, 1361 Route des
Lucioles, 06500 Valbonne, France}}
\maketitle

\begin{abstract}
We introduce a dissipative
 version of the Zhang's model of Self-Organized Criticality,
where a parameter allows to tune the local energy dissipation.
We analyze the main dynamical features of the model 
and relate in particular the Lyapunov spectrum with 
the transport properties in the stationary regime.
We develop a thermodynamic
formalism  where we define formal  Gibbs
measure, partition function and pressure
 characterizing the avalanche distributions.
We discuss the  infinite size limit in this setting.
We show in particular that a Lee-Yang phenomenon occurs in this model, 
for the only conservative case. This suggests
 new connexions to classical critical phenomena.
\end{abstract}

\bigskip


\bigskip

\bigskip

\bigskip

\baselineskip18pt

In 1988, Bak, Tang and Wiesenfeld (BTW)
\cite{BTW} proposed
for the first time a mechanism allowing a dynamical
system to reach ``spontaneously'' a steady state exhibiting some features
of a thermodynamic system at a critical point. Namely, 
by its only
internal reorganization in reaction  to  (stationary)
 external world perturbations, the system reaches a stationary
state with  power law statistics. This effect, called by these
authors Self-Organized Criticality (SOC), was quite
unexpected since attaining the critical state of a thermodynamic
system usually needs a fine tuning of some control parameter (temperature,
 magnetic
field, etc...)
that is at first sight absent from the definition of the BTW model
and its many variants \cite{Bak1,Jensen}.
This is a first difference between the SOC systems and 
thermodynamic systems exhibiting a second order phase transition.
Furthermore, at stationarity,
there is a constant flux
 through the system since the stationary flux of external perturbations  is dissipated 
in the bulk or at the boundaries. This corresponds
to a non-equilibrium situation and  it is therefore believed that usual
equilibrium statistical mechanics treatments using the concept
of Gibbs measure, free energy, etc ...
 cannot be applied in this case. 
Moreover, there is an implicit  notion of 
``thermodynamic limit'' in SOC systems in order to reach the critical state. 
However, the absence of a thermodynamic setting raises difficulties
in establishing a proper definition of the thermodynamic limit 
and only a few paper (like \cite{Maes})  
have focused on this aspect. 
Consequently, 
the results concerning this limit rely mainly on extrapolations
from the finite size numerical data and the properties  of the  limiting system
 have to be
conjectured from the finite size dynamical system. Unfortunately, these numerical
results must be handled properly. This implies in particular to have the right finite
size scaling form   \textit{and} a fair control
of the \textit{bias} induced by numerics.

In \cite{BCK1,BCK2,BCK3,BCK4} we  proposed an analysis of one SOC model,
the Zhang's model \cite{Zhang}, using the tools and concepts from dynamical
system theory and ergodic theory. In particular, our results opened the perspective
to construct a thermodynamic formalism in the sense of Sinai, Ruelle, Bowen \cite{Sinai,Ruelle,Bowen}.
 In this setting, one  defines in particular an extension of
the notion of Gibbs measure, where the Hamiltonian is replaced by a dynamically
relevant quantity. We discussed in particular in \cite{BCK1,BCK3} the possibility
to use this formalism to relate the microscopic dynamics to the macroscopic characteristics
of the critical state. Actually, non trivial results relating the fractal structure
of the attractor and the probability distribution of avalanche sizes were obtained.  
Moreover, we argued that this formalism could be used to define properly the stationary state
of the finite size dynamical system, and then an extrapolation of its properties
when the size tends to infinity could be done.

This paper and \cite{CM} are intending to develop this aspect. More precisely,
our aims are:

\bit

\item Construct the equivalent of finite volume Gibbs measures (in the
sense of Sinai, Ruelle, Bowen) in order to have
the equivalent of  the statistical mechanics generating functions :
partition function and free energy, to characterize the stationary state.
 Since
there is no Hamiltonian in the dynamics this construction must be done
in a more general setting.
 
\item Link these generating functions to dynamical properties.
 
\item Relate these functions to the avalanche distributions.
 
\item Analyze the behavior of these functions for finite size and attempt
to characterize the SOC state by extrapolating to the limit.
 
\item Have a theoretical control of the validity of the extrapolations
made from numerical simulations.
\eit

The two last points are the main topic of a separated and independent paper \cite{CM}.
In the present paper we will mainly focus on the three first aspects.
For, we introduce a non conservative
version of the Zhang model where a parameter 
$h$ controls the local energy dissipation. This parameter acts
somehow as a temperature and we show that it has to be tuned to $h=0$
in order to obtain a critical state in the thermodynamic limit.
Consequently, our model ``self-organizes'' into a ``critical'' state
only in the conservative case. However, the presence of this additional
parameter gives us some additional freedom to control the dynamics,
and allows us to define an additional critical exponent. Moreover, it allows
us to outline the role of boundary dissipation in the process
of self-organization into a critical state.

The paper is structured as follows.
In the first section we analyze in details the general properties
of the dynamical system such as
transport  and links to  Lyapunov exponents. The second
section is devoted to the construction of a
 thermodynamic
formalism  allowing us to define formal Gibbs
measure, ``partition function'' and ``free-energy''
 characterizing
 the stationary state.  We discuss the thermodynamic limit in this setting, in the third section.
We show in particular that, when the size of the system diverges,
 a \textit{Lee-Yang phenomenon}\cite{LY}
occurs, related to a loss of analyticity of 
 the formal free
energy. This unexpected result suggests that SOC systems might be closer as
previously believed to classical critical phenomena and opens some effective way
to ''map self-organized criticality to criticality'' \cite{Sornette}.  We show that the Lee-Yang 
phenomenon
is observed only when the energy is \textit{locally conserved} ($h=0$).
Moreover, the critical exponent characterizing the divergence of the correlation length of 
the avalanche size distribution when $h \to 0$ is analytically related to 
the angle that the Lee-Yang zeros do with the real axis with some analogy
with usual critical phenomena.
In section \ref{FDT} we also discuss the genericity of an important effect occuring
in the Zhang model: a weak form of initial conditions sensitivity.
We wrote an appendix where a short presentation of the thermodynamic formalism,
devoted to non-specialists, is presented.

\su{Dynamical properties of the finite-size model.}

\ssu{Definitions.}\label{Def}

The Zhang's model is defined as follows. 
Let $\Lambda$ be a d-dimensional box in $\bbbz^d$, taken as a
square of edge length $L$ for simplicity. Call $N=\#\Lambda=L^d$,
where $\#$ denotes the cardinality  of a set. 
Each site $i \in \Lambda$
is characterized by its
"energy" $X_i$, which is a non-negative real and finite number. 
Denote by $\bX  = \lbrace
X_i \rbrace_{_{i \in \Lambda}}$  a configuration of energies.
Let $E_c$ be a real, positive  number, called the 
\textit{critical energy},
and
$\cM = [0,E_c[^N$.
A configuration
$\bX$ is \textit{stable} when $\bX \in \cM$
and \textit{unstable} otherwise. 
In an unstable configuration the sites
 $i$ such that $X_i \geq E_c$ are called \textit{active}.
The dynamics action  on $\bX$ 
depends whether $\bX$ is stable
or unstable.

If $\bX$ is stable,
one chooses a site $i\in \Lambda$ at random with probability 
$\frac{1}{N}$, and add to it the energy
$\delta  = 1$ (\textit{excitation}).
If $\bX$ is unstable,
each active site  loses a part of its energy,
redistributed in equal parts to its $2d$ neighbors 
in the following way (\textit{relaxation}).
Fix two
\footnote{Note that the original Zhang's model corresponds to 
the case
$\epsilon=0$. The straightforward extension proposed
here allows us to avoid pathological dynamical
effects due to the existence of zero eigenvalues
for the tangent maps when $\epsilon=0$ \cite{BCK3}.} real parameters, 
$\epsilon \in [0,1[$
and $h \geq 0$. 
 Set $\gamma=\epsilon^h, \ \epsilon'=
\epsilon\gamma=\epsilon^{1+h}, \ 
\alpha=\frac{(1-\epsilon)}{2d}$. 
When $i$ is active 
 it gives the energy $\alpha X_i$ to its $2d$ neighbors
and keeps the energy $\epsilon'X_i$.
Therefore the energy is locally conserved  in the
case $h=0$ whereas there is local dissipation
\footnote{The case $h<0$ would correspond to local energy injection. However,
in this case there may not exist a stationary regime (see section \ref{Statio}).}
 when
$h>0$.
If several nodes are simultaneously active,
the local distribution rules are
additively superposed,
i.e. the time evolution of the system is synchronous.
It is useful to write down the relaxation dynamics with the map:

\beq \label{F}
\bF(\bX)=\bX + (\gamma-1)\epsilon \bZ(\bX)\ast\bX +\alpha\Delta\left[\bZ(\bX)\ast\bX \right]   
\eeq

In this equation $\bZ(\bX)$ is a $N$ dimensional vector,
such that $Z_i(\bX)=0$ if $X_i$ is stable
and $Z_i(\bX)=1$ if $X_i$ is unstable.
The $\ast$ denotes the product component by component,
that is if $\bX,\bY$ are $N$ dimensional vectors,
$\bX \ast \bY$ is the $N$ dimensional vector of components
$X_i Y_i$.  $\Delta$ is the discrete Laplacian on $\Lambda$.

The succession
of relaxations leading an unstable configuration
to a stable one is called an {\it avalanche}. 
Let $\dL$
be the boundary of $\Lambda$, namely the set of points
in $\bbbz^d \setminus\Lambda$
 at distance $1$ from $\Lambda$.
 The sites of $\dL$ have always
zero energy (dissipation at the boundaries). Since
the total energy in a finite lattice is finite,
all avalanches stop within a {\it finite} number of
iterations for $h \geq 0$. The addition of energy is {\it adiabatic}. 
When an avalanche occurs,
one waits until it stops before adding a new energy quantum.
Further excitations eventually generate a new avalanche,
but, because of the adiabatic rule, each new avalanche starts from {\it only one}
active site. 

The structure of an avalanche 
is encoded by the sequence of active sites
$\cC = \left\{C_t\right\}_{1 \leq t \leq \tau_\cC}$ where 
  $C_t = \left\{j \in \Lambda | X_j \geq E_c \ \mbox{in the t th step
of avalanche}\right\}$ and where $\tau_\cC$, called the \textit{avalanche
duration}, is the smallest
number such that $C_{\tau_\cC+1}=\emptyset$. The step $t=1$
corresponds to the initial excitation of some site $i$.
Consequently either $C_1=\emptyset$ (the addition of an energy quantum
$\delta$ is not sufficient to make the site $i$ active) or
$C_1=\left\{i\right\}$. Therefore,  
each avalanche can be labeled by a double index
$\left( i
,j \right),$ where the first index refers to the site
where the energy is dropped and the second one to an enumeration of the possible
avalanches starting at $i$ (including the ``empty'' avalanche where the
excitation of $i$ does not render it active).
The total energy of a stable configuration in a finite lattice being finite,
(it is bounded by $L^d E_c$), the total number of possible avalanches 
is finite for finite $L$,
(but diverges as $L \to \infty$ when $h=0$ as discussed below).
Call $l(i)<\infty$ the number of possible avalanches starting at $i$
(note that $l(i)$ depends on $L$ and $h$).
 The \textit{size} $s$ of an avalanche
is the total number of active sites, 
$s(\cC)=\#\cup_t C_t$. The \textit{area} $a$ is the number
of distinct active sites in $\cC$. 
Furthermore, one can use  $a,s,\tau$  to
make a partial ordering of the avalanches : $\cC \prec \cC' \Leftrightarrow
n(\cC) \leq n(\cC')$. The size $s$, area $a$, and duration $\tau$
are examples of avalanches observables. 
We call avalanches observables a function
$n$ which assign to an avalanche a positive real number and such that
$\cC \subset \cC' \Rightarrow n(\cC) \leq n(\cC')$. \\

 Because
all avalanches are {\it finite} (for finite $L$), and
since we are not interested in the  transients, one can, without loss of
generality take all initial energy configurations $\bX \in \cm$. All
 trajectories
starting from $\cM$ belong to a compact set $\cD$.
 Call $\bar{\cM}= \cD \setminus \cM$. $\bar{\cM}$ contains  the set of all
unstable
energy configurations achievable in an avalanche starting from an energy configuration in $\cM$.
It is useful to encode the dynamics of excitation in the following way.
 Let
$\Sigma^+_\Lambda$ be the set of right infinite sequences 
$\ta = \left\{ a_1,\dots, a_k, \dots\right\}, a_k \in \Lambda$, and $\sigma$
be the {\it left shift}\footnote{Throughout this paper we will often
 think of the left
Bernoulli shift on $\Sigma^+_\Lambda$ as represented by the
 system $z \to Nz \ \mbox{mod}\ 1, z \in [0,1]$. This allows in particular
 to
define the tangent map in the direction of the shift. \label{Nx}} over 
$\Sigma^+_\Lambda$, namely
$\sigma \ta = a_2a_3 \dots $. The elements of $\Sigma^+_\Lambda$ are called
{\it excitation sequences}. The set $\Sigma^+_\Lambda \times
\cD$ is the {\it phase space} of the Zhang's model
and $\hX=(\ta,\bX)$ is a point in
the phase space.
 The Zhang's model dynamics is given by a map $\hF : \Sigma^+_\Lambda \times
\cD \rightarrow \Sigma^+_\Lambda \times
\cD$ such that:
\bea\label{sdloc} 
\bX \in \cm \Rightarrow \hF(\hX) &=&
\left(\sigma\ta,\bX+\be_{a_1} \right)\\
\bX \in \bcm \Rightarrow \hF(\hX) &=& \left(\ta,\bF(\bX)\right) \eea

\nid where $\be_a$ is the $a$-th canonical basis vector of $\bbbr^N$.
The knowledge of an initial energy configuration $\bX$ and of an 
excitation sequence  $\ta$  fully determines the evolution.
One can endow $\Sigma^+_\Lambda$ with a probability distribution
corresponding
to  a random choice of the excited sites.
The excited sites are chosen at random and
independently with
 uniform probability. This corresponds to endowing $\Sigma^+_\Lambda$
with the \textit{uniform
Bernoulli measure}, denoted by $\nu_L$ in the sequel.\\

The model definition entails the convergence of the dynamics to
a stationary state where the average incoming energy flux is compensated
by the dissipated energy flux. It is furthermore conjectured in the
SOC literature that this state is  reached independently
of the initial state.  In the dynamical systems terminology
this amounts to  assuming that there exists a \textit{unique natural} or 
\textit{Sinai-Ruelle-Bowen}
(SRB) measure \cite{Sinai,Ruelle,Bowen}. Indeed, in this case, one has the following properties.
Let $\hm_L$ be an invariant measure on $\Omega$. Let $\psi$ be some function, 
(integrable with respect to $\hm_L$). 
Denote by $ \bar{\psi}_L$ the time average :

\beq
 \bar{\psi}_
L(\hX) \deq
\lim_{T \rightarrow \infty}\frac{1}{T}\sum_{t=1}^{T} \psi(\hF^{t}(\hX))
\eeq

\nid  and let $\int\psi(\hX)d\hm_L(\hX)$ be the ensemble average with respect to $\hml$.
Then $\hm_L$ is a SRB measure if $\bar{\psi}_L(\hX)
=\int\psi(\hX)d\hm_L(\hX)$ for a set of initial conditions
$\hX$ of \textit{positive Lebesgue measure} \cite{Keller}.
If the SRB  measure is unique then the time average
and the ensemble average with respect to $\hm_L$ are equal
for a set of initial conditions of \textit{full} Lebesgue measure.

Equivalently, in this case, the SRB measure is the weak limit :

\beq
\hml=\lim_{T\to \infty} \frac{1}{T} \sum_{t=1}^T \hF^{\ast t}(\nu_L \times \mu_L^{(0)})
\eeq

\nid for any absolutely continuous measure $\mu_L^{(0)}$ 
characterizing the distribution of the initial energy configurations,
where $\hF^{\ast t}(\nu_L \times \mu_L^{(0)})$
is the image of $\nu_L \times \mu_L^{(0)}$ under the map $\hF^t$. 
Note furthermore that if $\hF$ is topologically mixing \footnote{
For any open sets $\cO_1,\cO_2 \subset \cA$, $\exists 
T\equiv T(\cO_1,\cO_2) >0$ such that
$\cO_1 \cap \hF^t\cO_2 \neq \emptyset, \ \forall t \geq T$ \cite{Katok}}
then :

\beq \label{SRBmix}
\hml=\lim_{t\to \infty}  \hF^t(\nu_L \times \mu_L^{(0)})
\eeq

 The existence of a unique SRB measure is somehow the minimal property to 
assume
since it implies that, if one selects the initial conditions
 at random with a uniform probability (or any probability
having a density), then the time average will not depend on the initial condition.
Furthermore, the property (\ref{SRBmix}) corresponds exactly to what is numerically observed.
Starting from an arbitrary (absolutely continuous) probability measure $\mu_L^{(0)}$ on the
set of 
initial energy configurations 
and iterating the dynamics, the state $\hF^t(\nu_L \times \mu_L^{(0)})$, characterizing
the statistical distributions of points at time $t$, converges to the stationary
state $\hml$. This leads to state what we call the \textit{first SOC conjecture} in the sequel:

\bconj\label{SOCHyp}
For any finite $L$, their  exists a unique SRB  measure $\hml$ obeying (\ref{SRBmix}) 
for generic values of the parameters $E_c,\epsilon,h \geq 0$
and  whatever the lattice dimension $d <\infty$.
\econj

\nid Note that the projection of $\hm_L$ on $\Sigma^+_\Lambda$ is the 
Bernoulli measure $\nul$ by construction. 

Proving this property in the Zhang's model
 is clearly a difficult task which
is beyond the scope of this paper. We note however that this point
has been widely discussed in a previous paper \cite{BCK3},
where strong mathematical arguments in favor of this were given.
Actually, the existence of a unique SRB measure was proved, but
restricted to the one dimensional model,
and to some $E_c$ interval.
Note \textit{a contrario} that the failure
of this property would lead to the coexistence of several
 stationary states depending on initial
conditions.  This has not been observed.

Since the SRB measure is the stationary state, the physically relevant
quantities characterizing the stationary regime are obtained from
$\hml$.
Call $\boml\deq\hml(\Sigma^+_\Lambda\times\cM)$. By definition this is the probability
to inject energy in the system. The average incoming energy flux, at stationarity, 
is given by the vector  $\Bhl$  with components:

\beq
\Bhl(i)=\frac{\boml}{N} \deq \bhl
\eeq

\nid  corresponding to the average energy received
by the site $i$, per unit time. Note that in this paper we consider, for simplicity,
 the case where the energy is uniformly distributed in the lattice, though most of the results
hold in the more general case where $\Bhl$ is not spatially uniform.
The quantity $\bhl = \frac{1}{N}\sum_{i=1}^N \Bhl(i)$ is called the \textit{injection rate}.
This is the average energy injected per site and per unit time.

Define

\beq \label{defrhoL}
\rho_L(i)=Prob[X_i(t)\geq E_c]\deq 
\hml\left[\left\{\hX, \ X_i \geq E_c\right\}\right]
\eeq

\nid the probability that the site $i$ is critical in the stationary
regime, and let $\rho_L$ be the  $N$ dimensional vector,
$\rho_L=\left\{\rho_L(i)\right\}_{i=1}^N$. The quantity

\beq \label{defrhoLav}
\rho_L^{av}=\frac{1}{N}\sum_{i=1}^N\rho_L(i)
\eeq

\nid is often
called the \textit{density of active site} in the literature.
It has been introduced by Vespignani et al. \cite{Vespignani}
as an order parameter in SOC models.  Moreover, these authors introduced the quantity 

\beq \label{defsusceptibilite}
\chi_L(h) \deq  \frac{\partial \rho_L^{av}}{\partial \bhl} 
\eeq

\nid that they interpret as a 
 \textit{response
function} with respect to  variations in the injection rate.
In particular, considering the excitation at a given point and at a given
time as a perturbation in the relaxation dynamics, $\chi_L(h)$ 
characterizes the linear response  to this perturbation.
We
show below that, in our model, $\chi_L(h)$  \textit{diverges} when $h=0$ and $L \to \infty$, 
as expected for a critical phenomenon.\\

In the conservative case $h=0$,
the numerical simulations report the following behavior \cite{Zhang}.
Fix some avalanche observable $n$.
Call  $P_L(n)$ the probability distribution of $n$ in the stationary state,
for a box $\Lambda$ of size  $L^d$. It is observed that $P_L(n)$
 has  a power law
shape over a finite range, with a cut-off corresponding to finite size effects. As $L$
increases the power law range increases. 
In the SOC literature it is assumed that
 the limit $L \to \infty$ of the model is well defined,
 and that 
the corresponding probability distribution for $n$, $P^\ast(n)$,
 is a power law. This is  the
\footnote{Note that the first (existence of a unique
stationary state) and second  (convergence to a limit
state with power law statistics, as $L\to\infty$) SOC conjecture
are not inherent to the Zhang model but are implicit in all SOC
models. The mathematical proof of the existence of a unique
stationary state for a finite lattice has only been done in a few cases (such as
the Abelian sandpile) \cite{Dhar},
while the proof of the second SOC conjecture remains
to be done for all the main SOC models.} \textit{second SOC conjecture}:

\bconj
As $L \to \infty$, for $h=0$, for each observable $n$, $P_L(n)$ converges to a power
law 
$$P^\ast(n)=\frac{K}{n^{\tau_n}}, \ n = n_0 \dots \infty$$ 
\econj

\nid where $n_0$ is model dependent. 
This entails the scale invariance of the corresponding stationary state  
which has therefore some common features with a critical state.
$\tau_n$ is called the {\it critical exponent} of the observable $n$.
It is commonly admitted in the SOC community
that
a classification of the models can be made by the knowledge
of their  critical exponents (``universality classes''). 

When $h \neq 0$, the energy balance changes since one has
an additional dissipation (resp. injection) term when $h>0$ 
(resp.
$h<0$). We show in the section \ref{Distrib} that, when $h > 0$,
the stationary state cannot converge to
a critical state in the limit $L \to \infty$. This regime 
is called \textit{subcritical}. When $h<0$ one easily shows that
there exists an $h$ value, $h_L <0$, such that a stationary regime
exists only if $h>h_L$. The corresponding regime is called
\textit{supercritical}. It exists only for finite $L$ since $h_L \to 0$
as $L \to \infty$ (see section \ref{Statio}).

\ssu{Transport properties.} \label{Transport}

We now discuss the properties of the Lyapunov exponents and their connection
with the energy injection and transport in the Zhang's model.
This section presents original results which are an extension of
the theory developed in \cite{BCK4}.
Recall
that the Lyapunov exponents are the
logarithm of the eigenvalues of the $N+1 \times N+1$ matrix 
$\lim_{t \to \infty}\left[\tilde{D\hF}^t_{\hX}D\hF^t_{\hX}\right]^{\frac{1}{2t}}$
where $\tilde{}$ denotes the transpose. Note that the Oseledec multiplicative
ergodic theorem \cite{Oseledec} insures that the limit exists almost-surely and is
a constant.
The particular structure 
of the dynamics (\ref{sdloc}) implies that $D\hF_{\hX}$ is block diagonal with
a $N$ dimensional block, $D\bF_{\bX}$, corresponding 
to the relaxation dynamics and a one dimensional block corresponding 
to the excitation dynamics (see footnote \ref{Nx}). 

The excitation dynamics is expansive and has a positive 
Lyapunov exponent :

\beq \label{lambda0}
\lambda_L(0)=\bom_L\log(N)
\eeq

\nid Note that $\lambda_L(0)$ is also the
Kolmogorov-Sinai entropy.

The Jacobian matrix $DF_{\bX}$ of $\bF$ 
 characterizes the energy
transport during the avalanches, since
the entry  $DF^t_{\bX,ij}$  is the ratio of energy
flowing  from site $j$ to site $i$, in $t$ times steps, for the initial condition
 $\bX$  \cite{BCK4}. Moreover, the relaxation dynamics is contracting\cite{BCK3} and, consequently, 
the $N$ corresponding
Lyapunov exponents are \textit{strictly} negative.  Order them such that
 $0>\lambda_L(1) \geq \lambda_L(2) \geq \dots \geq \lambda_L(N)$.
This hierarchy of
Lyapunov exponents corresponds to a hierarchy of time scales
for the energy transport in the lattice.
Define $M(\bX,t)=\tilde{D\bF}^t_{\bX}.D\bF^t_{\bX}$ and
$\Lambda =\lim_{t\to \infty}
M(\bX,t)^{\frac{1}{2t}}$.
$M(\bX,t)$ being
symmetric it admits an orthogonal basis
$\left\{\bv_i(\bX,t)\right\}_{i=1}^N$ and eigenvalues $\mu_i(\bX,t)$.
One can show  that $\lambda_L(i)=\lim_{t \to \infty} \frac{1}{2t}log(\mu_{i}(\bX,t))$.
Furthermore, each $\bv_i(\bX,t)$ converges exponentially
to a vector  $\bv_i(\hX)$ in $\bbbr^N$,
depending on $\hX$ \cite{Pollicott}.  We call these vectors
 the {\it Oseledec modes} for the trajectory of $\hX$. They
can be numerically obtained from the QR decomposition used in the
computation of the Lyapunov spectrum (see \cite{ER}).
The collection of Oseledec modes defines
 a hierarchy of nested subspaces (Oseledec splitting):
$$\bbbr^N = \cV_1(\hX) \supset \cV_2(\hX) \supset \dots \supset \cV_N(\hX)$$
\nid depending on ${\hX}$, where $\cV_k(\hX)=span\left\{\bv_k(\hX), \dots, \bv_N(\hX)\right\}$,
 and such that the  norm
of a perturbation $\bv \in \cV_i(\hX)
\setminus \cV_{i+1}(\hX)$
is given by :
\beq
\|D\bF^t_{\bX}.\bv \| = C(\hX,t) e^{\lambda_L(i)t}\|\bv\|
\eeq

\nid where $\lim_{t \to \infty} \frac{1}{t}logC(\hX,t)=0$ almost
surely. $\lambda_L(i)$
is therefore the  exponential rate of variation of $\|\bv\|$. 
 This exponent is closely related
to the energy dissipation rate (see below and in \cite{BCK4}).\\

It has been shown in \cite{BCK4} that the spectrum of
negative Lyapunov exponents
 can
be roughly split into two parts: one part (fast modes
and largest Lyapunov exponents
in absolute value) corresponds to stabilizing modes contracting fast
the dynamics onto the attractor; the other part (slow modes) corresponds
to transport modes (see Fig. \ref{graphLyap}). The slowest transport
modes
 can be computed
from a linear operator $\cL : \bbbr^N \to \bbbr^N$
which acts on a vector $\bv \in \bbbr^N$ as:

\beq \label{L}
\cL(\bv)=\bv+ 2(\gamma-1)\epsilon\rho_L\ast\bv + 2\alpha\Delta\left(\rho_L\ast\bv\right)
\eeq

 The  Lyapunov exponents are given by the singular
values of $\cL$. We have drawn fig. \ref{graphLyap} the Lyapunov spectrum
for $L=20,E_c=2.2,\epsilon=0.1,h=0.1,d=2$ and the singular values of (\ref{L}),
with (Fig. \ref{graphLyap}a) and without (Fig. \ref{graphLyap}b) boundaries 
(in this case $\Lambda$ is a $2$ dimensional torus).
One checks that the slowest modes are well approximated.

\bef
\bc
\begin{minipage}{6cm}
\epsfxsize=6cm
\epsfysize=6cm
\epsffile{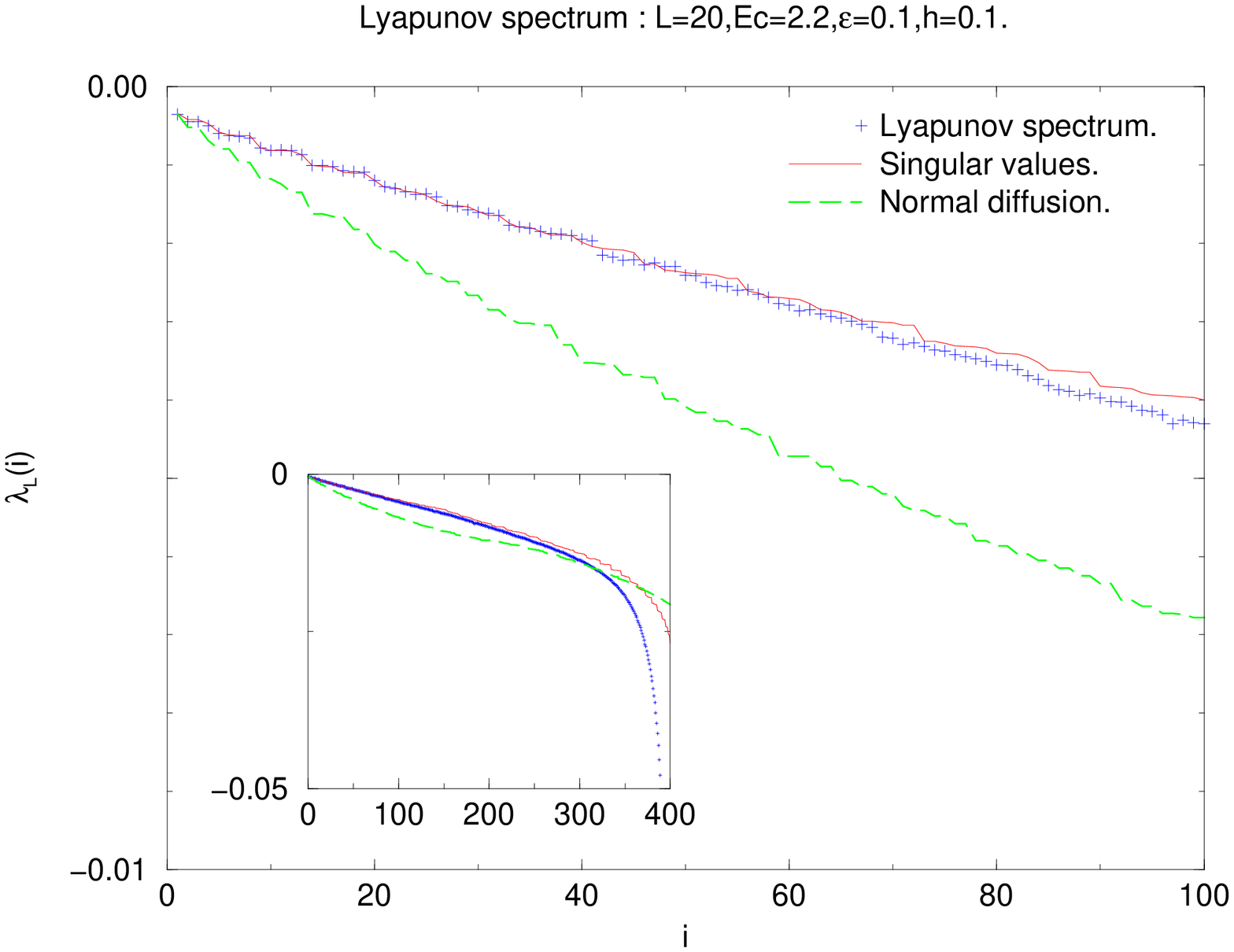}
\end{minipage}  \hspace{1cm}
\begin{minipage}{6cm}
\epsfxsize=6cm
\epsfysize=6cm
\epsffile{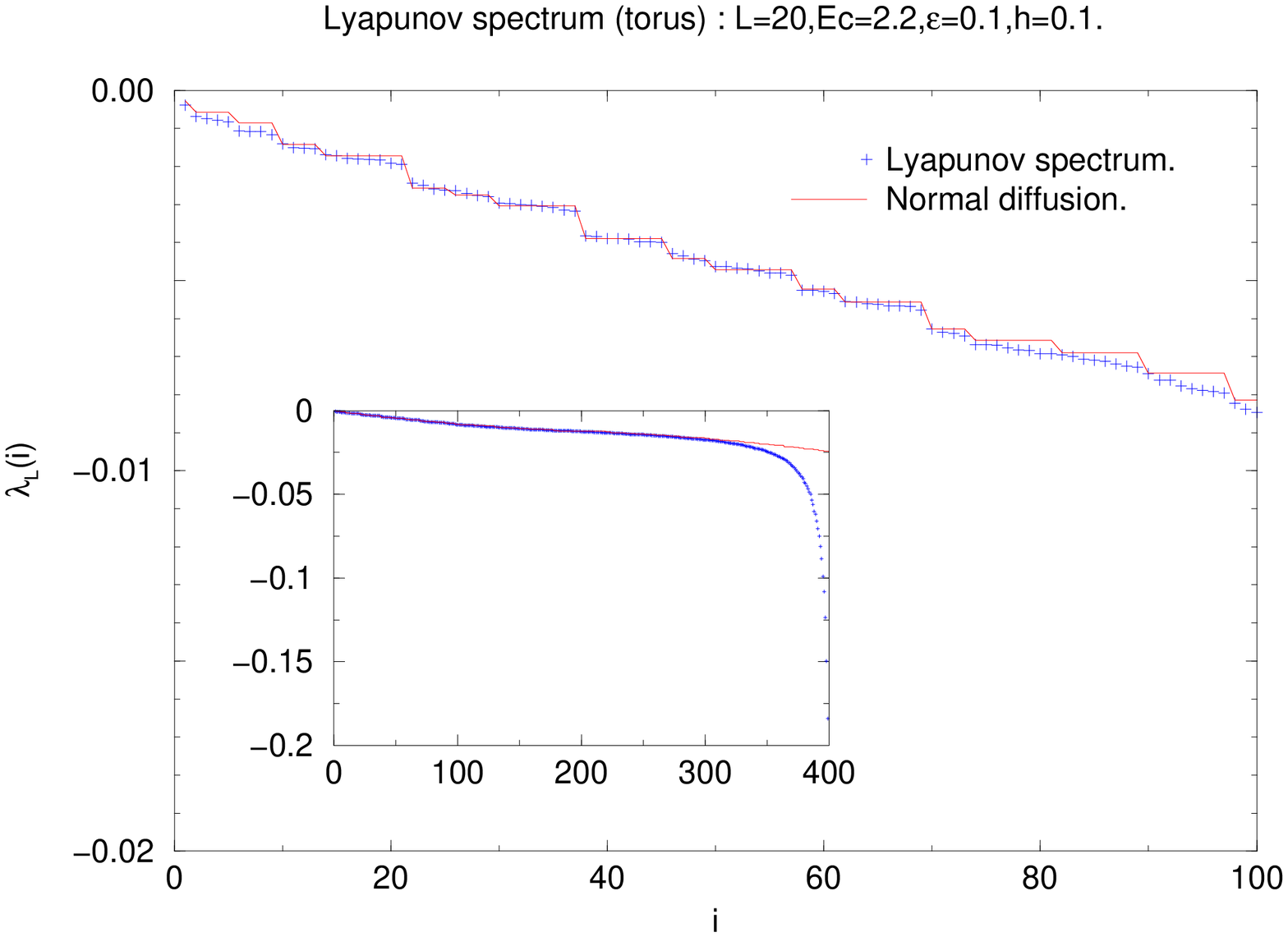}
\end{minipage} 
\caption{a. Lyapunov spectrum
for $L=20,E_c=2.2,\epsilon=0.1,h=0.1,d=2$ and the singular values of (\ref{L}).
Figure \ref{graphLyap} b. The Lyapunov spectrum for the same parameters values,
but without boundaries (torus).
\label{graphLyap}}
\ec
\enf 

$\cL$ can be interpreted as  the transfer  operator 
of a diffusion in a  metric given by $\rho_L$,
with a probability $2(\gamma-1)\epsilon\rho_L$ of trapping a particle
\footnote{The factor $2$ is due to the fact that a site cannot relax two successive
time steps (see \cite{BCK3,BCK4} for details).}. 
In the presence of boundaries 
 $\rho_L$ is non uniform in the lattice
(see fig. \ref{Figrho}a). The consequence is that the Lyapunov spectrum
departs from a diffusion in a flat metric, except for the first mode
(see Fig. \ref{graphLyap}a). A contrario, the same model defined on a torus
with bulk dissipation has a spatially uniform $\rho_L$.
Consequently, the Lyapunov spectrum is similar to the normal diffusion in a flat
metric, for characteristic times scales a bit larger than the average avalanche size. 
The Lyapunov exponents corresponding to smaller times scales departs from normal
diffusion, due to the presence of a threshold in the dynamics definition (see
fig. \ref{graphLyap} b).

This analogy with anomalous diffusive transport
is fruitful since it allows us to interpret the Lyapunov exponents and the corresponding
transport modes as the analogous of Fourier modes. In particular,
the largest negative Lyapunov, $\lambda_L(1)$, defines a characteristic time scale
$t_L(1)=\frac{1}{|\lambda_L(1)|}$ or \textit{microscopic escape rate},
 corresponding to the time taken by a particle
to exit the system, either by disappearing in the bulk, or by escaping from
the boundaries. In terms of the dynamical system (\ref{sdloc}) the largest
 negative Lyapunov
exponent is the exponential decay rate of a generic perturbation about
some point $\bX$. In particular the excitation
of the site $i$ corresponds to a perturbation about $\bX$ oriented in
the $i$ th canonical direction in $\bbbr^n$ and belongs generically
to $\cV_1(\hX) \setminus \cV_2(\hX)$. Consequently, $\lambda_L(1)$ gives 
the decay rate of the initial
local perturbation induced by the excitation and  $t_L(1)$ defines 
the characteristic time required for this perturbation to vanish.
 
The
 first Lyapunov exponent is well approximated by (Fig. \ref{firstLyap}) :

\beq \label{lambda1}
\lambda_L(1)  \sim -\frac{2\boml}{E^+_{tot}}
\eeq

\nid and scales therefore like the dissipated energy per unit time
(see  theorem 2 in  \cite{BCK4}).

\bef
\bc
\begin{minipage}{6cm}
\epsfxsize=6cm
\epsfysize=6cm
\epsffile{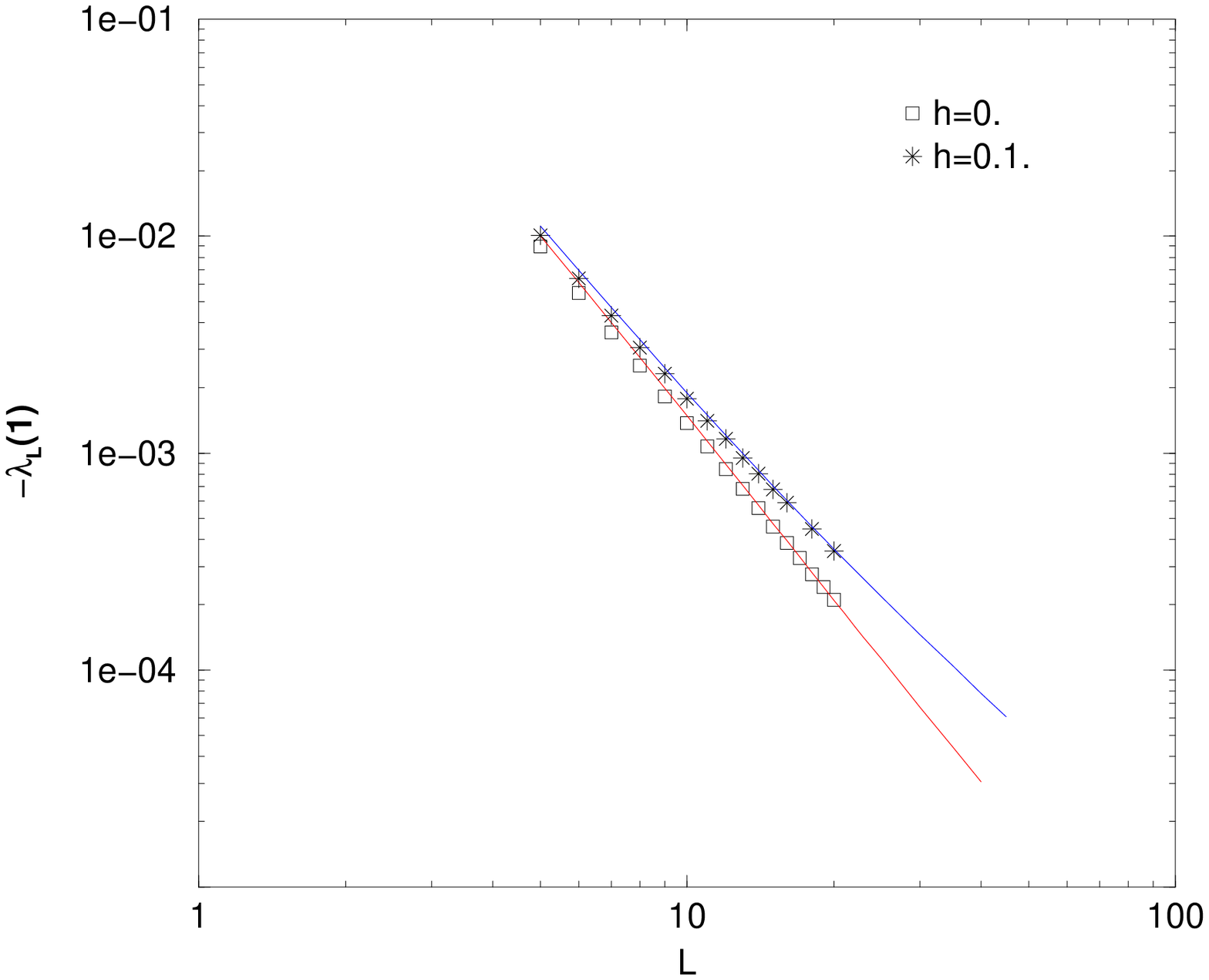}
\end{minipage}  \hspace{1cm}
\caption{First Lyapunov exponent $\lambda_L(1)$
versus $L$ for $E_c=2.2,\epsilon=0.1,h=0,h=0.1$. In full lines are drawn the 
theoretical values given by
(\ref{lambda1})
\label{firstLyap}}
\ec
\enf

\ssu{Stationarity  equations.} \label{Statio}

It is possible to obtain an approximate analytic expression
for $\rho_L$.
It can be  obtained 
by noting that :

\beq\label{balanceF}
\bar{\bF}_L -\bar{\bX}_L=
\int 
\left\{ (\gamma-1)\epsilon \bZ(\bX)\ast\bX +\alpha\Delta\left[\bZ(\bX)\ast\bX \right]
\right\}
d\hml(\hX)   
\eeq

\nid is the average energy lost by each site within one step
of the dynamics, in the stationary regime. At stationarity,
this loss is compensated by the average incoming flux
$\Bhl$. The balance equation writes:

\beq\label{Eq}
(\gamma-1)\epsilon \bar{\bU} +\alpha\Delta\bar{\bU} = -\Bhl
\eeq

\nid where $\bar{\bU} \deq \rho_L\ast\bar{\bX}^+_L$.
The vector $\bar{\bX}^+_L = \left\{\bar{X}^+_L(i)\right\}_{i=1}^N$
is such that the $i$ th component, $ \bar{X}^+_L(i)
\deq E\left[X_i | X_i \geq E_c \right]$,
is the \textit{conditional
expectation} of $X_i$ 
given that $X_i \geq E_c$. (Alternatively,
$\bar{X}^+_L(i)$ is the average energy of $X_i$ given
that $X_i$ is active).

It is easy to solve (\ref{balanceF}) by decomposing $\bar{\bU}$ on the
eigenmodes of the Laplacian. $\rho_L$ can then be obtained
by noting that the spatial fluctuations of $\bX^+_L$ 
are small (except near the boundary).
One therefore considers $\bX^+_L$ as spatially constant. This yields 
the following  ``first order'' approximation for $\rho_L$ :

\beq \label{rhoL}
\rho_L(\bx) = \sum_{\bn} A_{\bn} \prod_{i=1}^dsin(k_i x_i)
\eeq

\nid where  $\bx=(x_1, \dots, x_n)$ is the set of coordinates of a point
in a $d$ dimensional lattice,
$\bn=(n_1,\dots, n_d)$ is the set of quantum numbers
parameterizing the eigenmodes of the discrete Laplace operator,
 and
\beq
s_{\bn}=\left[(\gamma-1)\epsilon+
2\alpha(\sum_{i=1}^dcos(k_i)-d)\right]
\eeq
\nid is the corresponding  eigenvalue
with $k_i=\frac{n_i\pi}{L+1}$.

The coefficients $A_\bn$ are given by:

$$A_{\bn}=-\frac{2^{d}\bom_L}{E^+_{tot}(L+1)^d}
\frac{\prod_{i=1}^dC_{n_i}}
{s_{\bn}}$$

\nid where $E^+_{tot}= \sum_{i=1}^N \bar{X}^+_L(i)$ and,

\beq \label{Cni}
C_{n_i}=\sum_{x=1}^L sin(k_i.x)
=(-1)^{m_i}\frac{sin(\frac{n_i\pi L}{2(L+1)})}{sin(\frac{n_i\pi}{2(L+1)})}
\eeq
 
\nid where $n_i=2m_i+1$.

The density of active site $\rho_L^{av}=\frac{1}{N}\sum_{i=1}^N\rho_L(i)$ 
 is given
by:

\beq\label{rhoLav}
\rho_L^{av} =-\frac{\bom_L}{L^dE^+_{tot}}\gamma_L
\eeq

\nid where:

\beq \label{gammaL}
\gamma_L = \left(\frac{2}{L+1}\right)^d
\sum_{\bn}\frac{\prod_{i=1}^dC^2_{n_i}}
{s_{\bn}}
\eeq 

The equations (\ref{rhoL}) and (\ref{rhoLav}) are in very good agreement with the 
empirical values. We plot  Fig. \ref{Figrho}a the empirical and theoretical
$\rho_L$ for $L=40,Ec=2.2,\epsilon=0.1,h=0.1$ and
Fig. \ref{Figrho}b  the empirical and theoretical curve
$\rho_L^{av}$ as a function of $L$, for $Ec=2.2,\epsilon=0.1$  $h=0$ and $h=0.1$.

\bef
\bc
\begin{minipage}{8cm}
\epsfxsize=8cm
\epsfysize=8cm
\epsffile{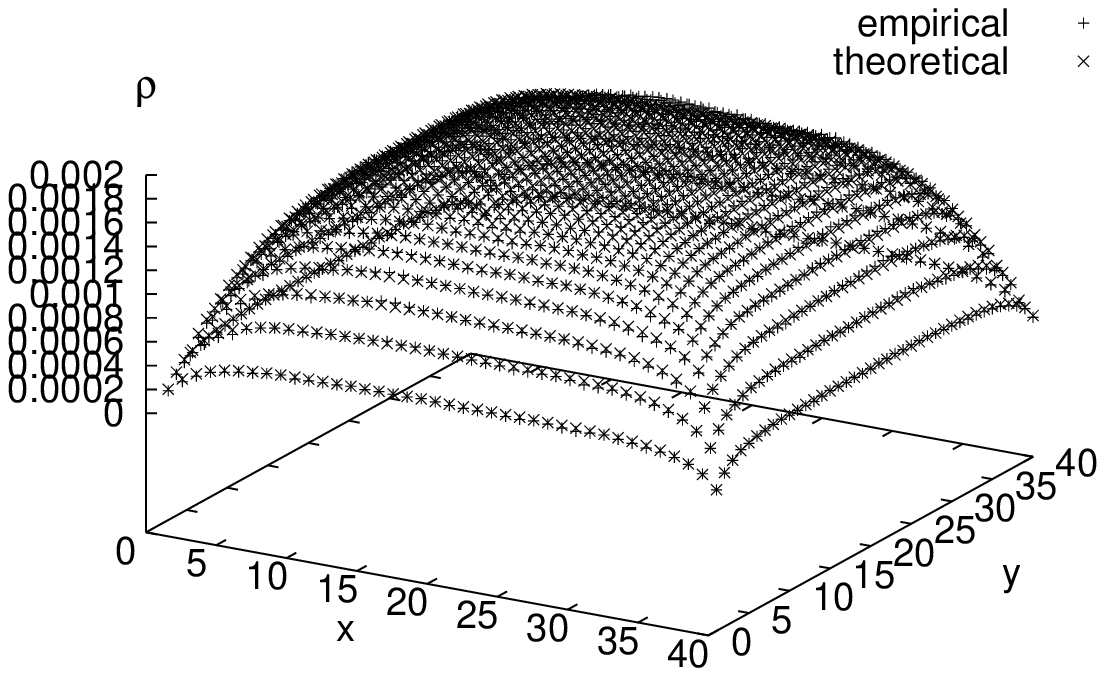}
\end{minipage}  \hspace{1cm}
\begin{minipage}{6cm}
\epsfxsize=6cm
\epsfysize=6cm
\epsffile{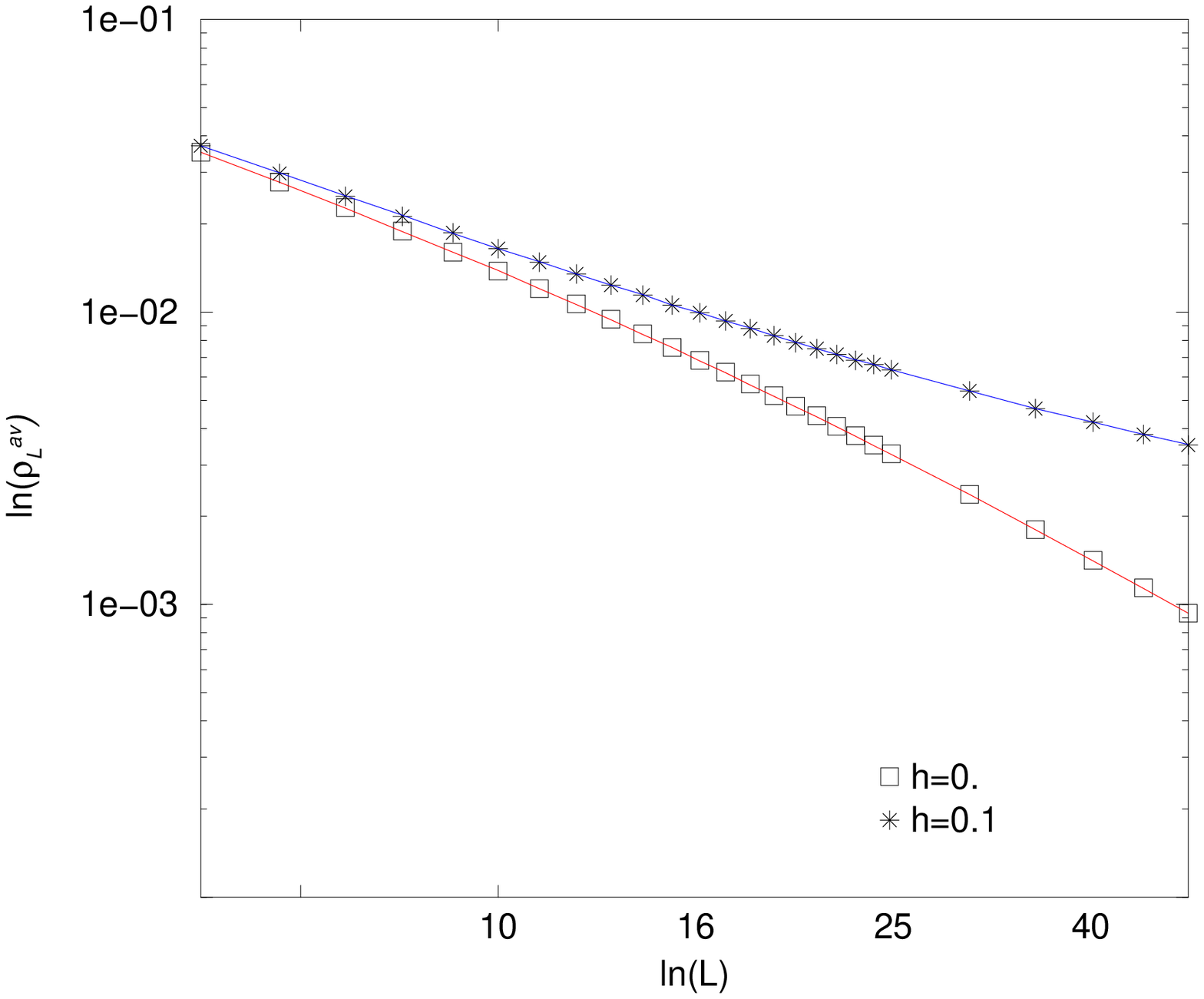}
\end{minipage}
\caption{ a. Empirical and theoretical
$\rho_L$ for $L=40,Ec=2.2,\epsilon=0.1,h=0.1$.
Fig. \ref{Figrho}b  the empirical and theoretical (full lines) curve
$\rho_L^{av}$ as a function of $L$, for $Ec=2.2,\epsilon=0.1$  $h=0$ and $h=0.1$.
\label{Figrho}}
\ec
\enf

The equation (\ref{rhoLav}) is a balance equation. It can indeed
be rewritten:

\beq \label{balance}
\rho_L^{av}\bar{e}_L= \bhl
\eeq

\nid where :

\beq \label{disrate}
\bar{e}_L=\frac{E^+_{tot}}{|\gamma_L|}
\eeq

\nid corresponds (from eq. \ref{balance}) 
to the average energy dissipated per unit time
and per active site. It is called the 
\textit{dissipation rate} in the literature \cite{Vespignani}.

Equation (\ref{balance}) implies that the susceptibility (\ref{defsusceptibilite})
obeys:

\beq \label{susceptibilite}
\chi_L(h) 
= \frac{1}{\bar{e}_L}-\frac{\bhl}{\bar{e}_L^2}\frac{\partial \bar{e}_L}{\partial \bhl}
=\frac{1}{\bar{e}_L}\left[1-\rho_L^{av}\frac{\partial \bar{e}_L}{\partial \bhl}\right]
\eeq

Were the excitation rate per site
$\bhl$ and the dissipation rate
$\bar{e}_L$ independent, would this relation 
reduce to $\chi_L(h) = \frac{1}{\bar{e}_L}$.
This corresponds to the result found  by Vespignani et al. 
in a sandpile model where the excitation rate per site and
dissipation rate were independent tunable parameters \cite{Vespignani}. 
 The
relation (\ref{susceptibilite}) is an extension of this result
to the case where $\bar{e}_L,\bhl$ are not independent and tunable parameters
but are determined by the microscopic evolution. (This is exactly what is
meant by ``self-organized'').\\

 As $L \to \infty$ it is easy to see
that the leading contribution in
(\ref{gammaL}) corresponds to modes $\bn=(n_1, \dots, n_d)$ such that
$n_k <\beta(L+1)$ for some $\beta >0$ that can be taken arbitrary small.
Indeed, for these modes, according to eq. (\ref{Cni}), (\ref{eL})
$C_{n_i} \sim \frac{2(L+1)}{n_i\pi}$ while 
$s_{\bn} \sim 
(\gamma-1)\epsilon-\frac{\pi^2}{(L+1)^2}\sum_{i=1}^dn_i^2$. 
Consequently, the sum on  these modes scales like $(L+1)^{2d} \sum^\ast 
\frac{\left(\frac{2}{\pi}\right)^{2d}}{n_i^{2d}\left[(\gamma-1)\epsilon
 - \alpha \left(\frac{n_i\pi}{(L+1)}\right)^2\right]}$ where $\sum^\ast$ denotes the sum 
on the  $\bn$'s such that $\sup_{k}n_k<\beta(L+1)$.
On the other hand, one can decompose the  remaining terms in the sum defining $\gamma_L$
into sums on the modes $\bn$ such that exactly $r$ components are smaller than $\beta(L+1)$.
Each sum scales like $(L+1)^{2r}$ where $r<d$. Therefore, as $L \to \infty$,
$|\gamma_L|$ scales like $a\frac{(L+1)^{d}}{2(1-\gamma)\epsilon}$ when $h>0$,
and like $c(L+1)^{d-2}$ when $h=0$, where $a,c$ are some non negative constants.
Furthermore, $E^+_{tot} \sim L^d$ since $Ec \leq \bpxl(i)<K, \ i=1 \dots N$, where $K$ is some constant
independent of $L$. Therefore,
when $L \to \infty$ the dissipation rate, $\bel$, obeys the scaling relation :

\beq\label{eL}
\bel \sim \bpxl\left[2(1-\gamma)\epsilon A+CL^{-2}\right]
\eeq

\nid where $A,C$ are some positive constants.
Consequently, when $h>0$, the dissipation rate
\textit{converges to a positive value as} $L \to \infty$,
while it converges to $0$ like $L^{-2}$ when $h=0$
In Fig. \ref{FigeL}a  we plotted the dissipation rate 
$\bel$. In full lines are drawn the fitting curves
$\bpxl CL^{-2}$
 (conservative case $h=0$) and $\bpxl\left[a+CL^{-2}\right]$ (non
conservative case $h=0.1$) corresponding to eq. (\ref{eL}),
where $A,C$ have been obtained by fitting.
We found $a=0.0238 \pm 0.0001$ and
$C=5.959 \pm 0.14$. In Fig. \ref{FigeL}b  $\boml$ is
 plotted .

\bef
\bc
\begin{minipage}{6cm}
\epsfxsize=6cm
\epsfysize=6cm
\epsffile{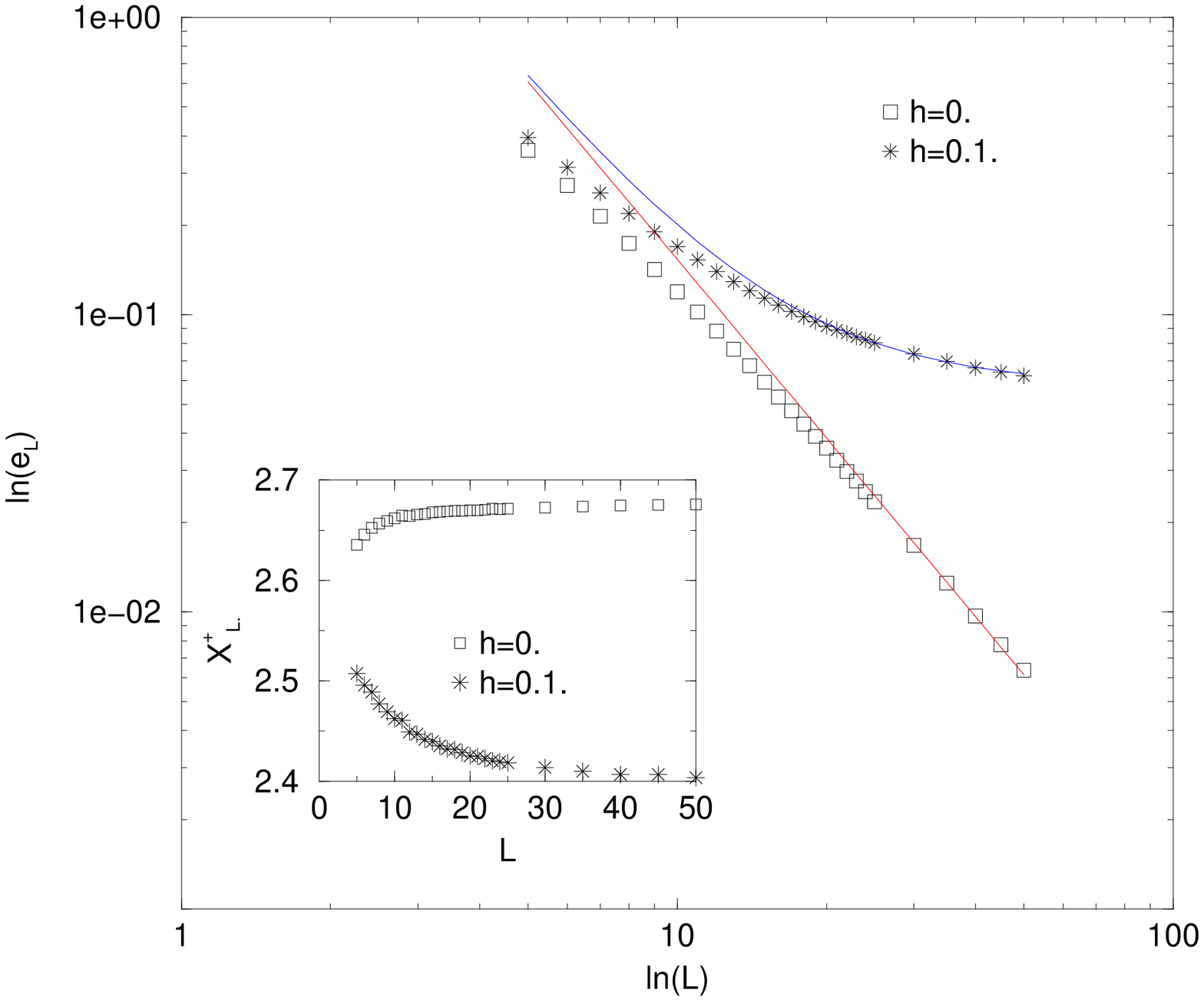}
\end{minipage}  \hspace{1cm}
\begin{minipage}{6cm}
\epsfxsize=6cm
\epsfysize=6cm
\epsffile{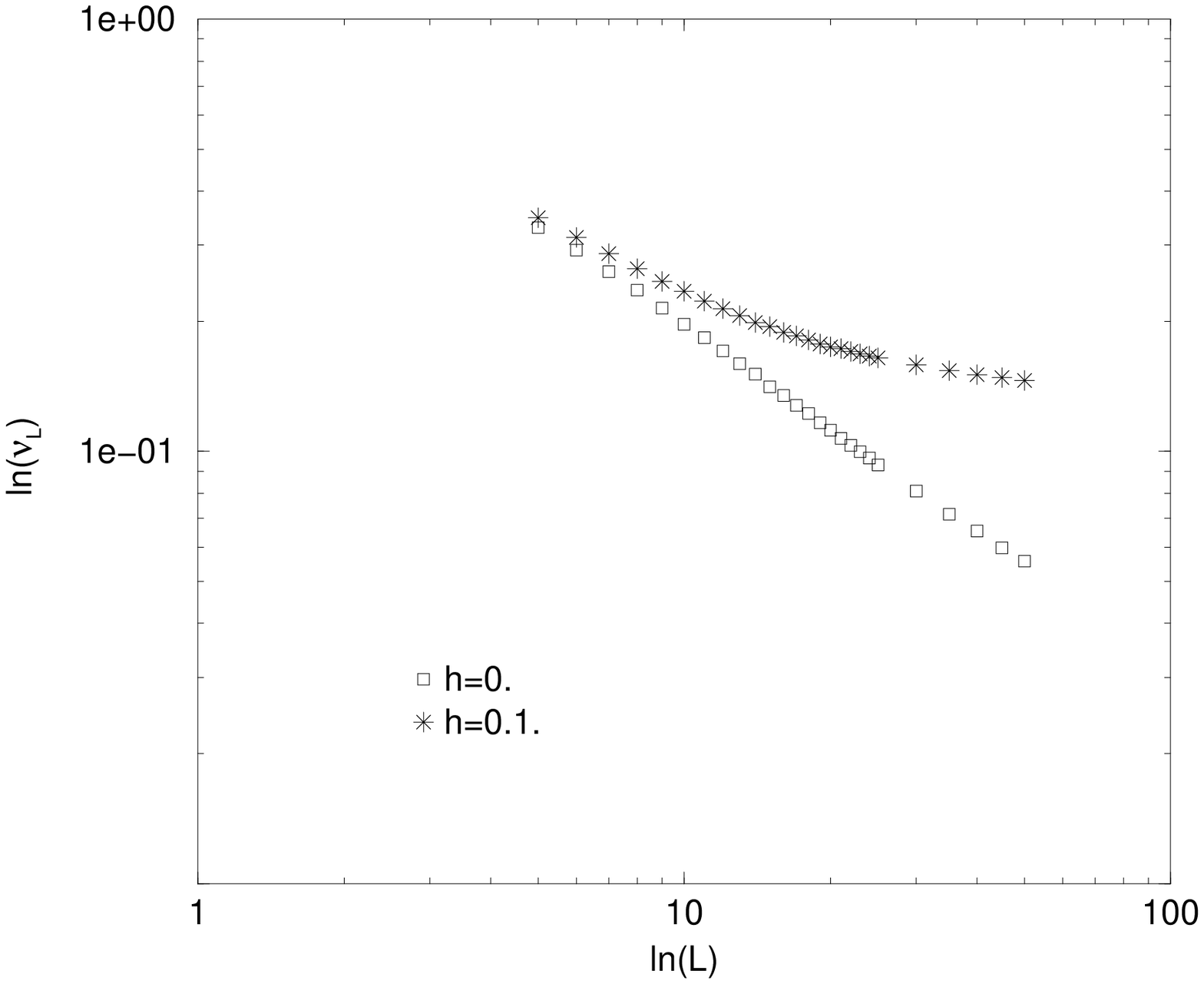}
\end{minipage}
\caption{ a Dissipation rate versus $L$ for $Ec=2.2,\epsilon=0.1$
in the case $h=0, \ h=0.1$ (insert: corresponding values of $\bpxl$).
In full lines are drawn the fitting curves
$\bpxl CL^{-2}$
 (conservative case $h=0$) and $\bpxl\left[a+CL^{-2}\right]$ (non
conservative case $h=0.1$) corresponding to eq. (\ref{eL}),
where $A,C$ have been obtained by fitting.
Fig. \ref{FigeL}b  The corresponding graph of $\boml$.
\label{FigeL}}
\ec
\enf

Note that  there exists a stationary
regime only if the dissipation rate is positive. Henceforth,
there exists a $h_L$ given by 
$\sum_{\bn}\frac{\prod_{i=1}^dC^2_{n_i}}{s_{\bn}}=0$
 such that, for $h<h_L$ there is no stationary
regime. $h_L$ behaves asymptotically like $\frac{
\log\left(1-\frac{CL^{-2}}{2\epsilon}\right)}
{\log(\epsilon)}$ and converges to $0$ as $L \to \infty$.

\ssu{Avalanches observable distributions.} \label{Distrib}

Call $P_L(n,h)$ the probability distribution of an avalanche observable
$n$, in the stationary regime, for a lattice of size $L$.
Call 

\beq \label{nL}
<n>_{L,h}=\sum_{n=0}^{\xi_L^n(h)}nP_L(n,h)
\eeq

\nid the  average value of $n$. $\xi_L^n(h)$
  is the maximal value that the random variable $n$
can take, among all the avalanches having a non
zero probability of occurrence in the stationary regime, 
in a lattice of size $L$. Since the energy used to initiate an avalanche is dropped locally,
$\xi_L^n(h)$ is the largest scale where the effect of a local perturbation
is observed, within one avalanche. 
In this way it corresponds to a correlation length \textit{within one avalanche}. 
This is a function of $h$. As discussed above $\xi_L^n(h) < \infty$ for $h \geq 0$, $L < \infty$.
To simplify the notation we will set : $P_L(n,0) \equiv P_L(n)$ (resp. $<n>_{L,0}
\equiv <n>_{L}, \  \xi_L^n(0) \equiv \xi_L^n$).

In the subcritical regime $h >0$  the avalanche size (resp. duration, area, etc..)
is bounded $\forall L$. Indeed, consider the evolution, under the singular diffusion (\ref{F}),
of a configuration $\bY$ in the \textit{infinite} 
lattice\footnote{With bulk dissipation since $h>0$} $\bbbz^d$, such that
 $E_c >Y_i \geq E_c - \eta$, $\eta>0$, $\forall i \in \bbbz^d$, with excitation
at some point $i_0 \in \bbbz^d$.
Consider simultaneously the normal damped diffusion
$\bY'(t+1) = \bY'(t) + (\gamma-1)\epsilon \bY'(t) +\alpha\Delta\left[\bY'(t) \right]$ 
on $\bbbz^d$ with a source $\delta=1$ applied at time $t=0$ at the site $i_0$,
where $Y'_i(t)=max(Y_i(t)-(E_c-\eta),0)$. By definition 
$Y_i(t)\leq Y'_i(t)+E_c-\eta$, $\forall t \geq 0$. $\bY'(t)$ converges asymptotically
to $0$ and it is easy
to show\footnote{On $\bbbz^d$ the eigenmodes of the damped
diffusion equation are $\lambda(k)=(\gamma-1)\epsilon - \alpha k^2$ where
$k$ is the wave vector in the Fourier space. The presence
of the damping coefficient $(\gamma-1)\epsilon$ 
insures the existence of the region $\cR$.} that there exists a bounded region  $\cR \subset \bbbz^d$
containing $i_0$, and a time $t_0$
 such that, $\forall t \geq t_0$ and $\forall i \in \bbbz^d \setminus\cR$,
$0 \leq Y'_i(t) < \frac{\eta}{2}$. Since $Y_i(t)\leq Y'_i(t)+E_c-\eta$,
$Y_i(t)\leq E_c-\frac{\eta}{2} < E_c$, $\forall t \geq t_0$, $\forall i 
\in \bbbz^d \setminus\cR$.
Consequently, the avalanche $\cC(\bY)$ does not got beyond
$\cR$ and the duration (resp. area, size) of $\cC(\bY)$ is bounded. Now,
since the largest avalanche size in a finite box $\Lambda \subset \bbbz^d$
is generated from a stable configuration $\bX$ such that
$X_i < E_c$ it is possible to find a configuration $\bY$ as above such
that $\cC(\bX) \subset \cC(\bY)$, provided $\eta$ is sufficiently
small (take $\eta <E_c-max_{i \in \Lambda}X_i$). Consequently, $\xi_L^n(h) < n(\cC(\bY)) <\infty,
 \forall L, \forall h >0$.
 It follows that the maximal size, area, etc...  are bounded
together with the corresponding average values when $h >0$. 
 Moreover, the energy excitation rate, $\boml(h)$,
scales like $\frac{1}{<\tau>_{L,h}}$. Consequently, $\boml(h)$ converges
to a \textit{strictly positive} value in the subcritical regime
(see Fig \ref{FigeL}b).

On the other hand, the average avalanche observable and the corresponding 
correlation lengths  \textit{diverge} in the conservative case, when $L \to \infty$.
Firstly, the behavior of the average avalanche size  $<s>_{L,h}$
as a function of $h$ and $L$ is obtained from 
a stationarity condition insuring the balance
between the incoming energy flux ($\delta=1$ is injected each time an 
avalanche
ends) and the average energy flux dissipated within an avalanche 
($<s>_{L,h}\bel(h)$). This gives:

\beq \label{scals}
<s>_{L,h} \sim  \frac{1}{\bel(h)} \sim \frac{1}{\bpxl\left[2(1-\gamma)\epsilon A+CL^{-2}\right]}
\eeq

\nid Therefore, $<s>_{L,h}$ converges
to a constant when $h>0$, while it \textit{diverges} like $L^2$
in the conservative case \cite{BCK4}. The adiabaticity condition
imposes that the number of active site
at the $t$-th avalanche step is bounded from above\footnote{This is
certainly not an optimal bound since, in particular,  a site cannot relax two successive
time steps.} by $\sum_{k=1}^{t}(2d)^{k-1}$.
This implies that the expectation of the
 avalanche
observables $a,\tau$ also diverges. It is observed  that  $<n>_L$ diverges
like:

\beq \label{divnL}
<n>_L \sim L^{\gamma_n}
\eeq

\nid where $n=a,s,t$, and $\gamma_n>0$
 in the conservative case.

Since most of the results available (in particular the numerical ones) are
obtained for finite $L$, one has to propose a finite size scaling form
allowing to extrapolate to $L \to \infty$ from finite size lattices results.
The most common scaling form
(see in particular \cite{Vespignani})
is  the following  (adapted to our notations):

\beq \label{PLnh}
P_L(n,h)=n^{-\tau_n}f(\frac{n}{\xi_L^n(h)})\ , \ n > n_0
\eeq

\nid where $n_0>0$ is model dependent.
 
This scaling form is discussed in more details in section \ref{Gibbs}, but 
 this is obviously the analog of the finite size scaling forms used
in  critical phenomena. The form (\ref{PLnh}) implies  $<n>_{L,h} \sim 
{\xi_L^n(h)}^{(2-\tau_n)}$.
Therefore, in the conservative case, $\xi_L^n(0) \sim L^{\beta_n}$,
where:
$$\beta_n=\frac{\gamma_n}{2-\tau_n}$$

\nid and  $P_L(n) = n^{-\tau_n}f(\frac{n}{L^{\beta_n}})$. 
This corresponds
to the Kadanoff et al. \cite{Kadanoff} finite size scaling form (\ref{Kada}).
Clearly, in this scaling form, $\tau_n,\beta_n$ are characteristics exponents 
allowing to determine the universality class of $P^\ast(n)$.

The form (\ref{PLnh}) also implies
that the correlation length of
\textit{avalanche size} is a function of the dissipation rate 
with a scaling form :

\beq \label{xiL}
\xi_L^s(h)=\bel(h)^{-\frac{1}{\sigma}} 
\eeq

\nid where $\sigma=2-\tau_s$.
Therefore, for $h=0$, the correlation length diverges
like $L^{\beta_s}$ where $\beta_s=\frac{2}{2-\tau_s}$,
and, for $h > 0$, $\xi_L^s(h)$, converges to a constant
$\xi^s(h) =\left(2A(1-\gamma)\epsilon\right)^{-\frac{1}{\sigma}}$.
Finally, as $h \to 0$, $\xi^s(h) \sim h^{-\delta}$, where 
$\delta=\frac{1}{\sigma}$.
Consequently, $\delta$ is a characteristic exponent  related to the 
singularity of  correlation length $\xi_L^s(h)$ as $h \to 0$.
In the section \ref{LeeYang} we show how
exponents $\tau_n,\beta_n,\sigma$ can be determined from the behavior of
the  zeros of a suitable partition function in full analogy
with usual critical phenomena .\\

From eq. (\ref{rhoLav},\ref{eL})
 the density of active site behaves asymptotically
like:

\beq \label{roLas}
\rho_L^{av} \sim \frac{\bom_L}
{E^+_{tot}\left[2(1-\gamma)\epsilon A +CL^{-2}\right]}
\eeq

Consequently, the density of active site converges
to $0$ whatever $h\geq0$. It scales like $\frac{1}{E^+_{tot}}
\sim L^{-d}$ for $h>0$ and like $\frac{\bom_L}{L^{d-2}}\sim
L^{2-d-\gamma_\tau}$
for $h=0$, where $<\tau>_L \sim L^{\gamma_\tau}$.

Finally, let us  discuss the behavior of the susceptibility 
(\ref{susceptibilite}).
It behaves like $\frac{1}{\bel}=<s>_L$ as $L \to \infty$ unless
\footnote{To avoid this case it is enough 
to assume that $\left|\frac{\partial \bar{e}_L}{\partial \bhl} \right| < 
\infty,
\ \forall L$ since $\rho_L^{av} \to 0$. As noticed above
$\bel, \bhl$ are not independent parameters and are constrained by the 
dynamics.
They depend on the stationary state which is itself determined by the control
parameters $E_c,\epsilon,\delta$. It is therefore sufficient to assume that
the variations of $\bel,\bhl$ are regular with respect to variations
in the control parameters. This holds unless the system is at a bifurcation point.}
 the term $\rho_L^{av}\frac{\partial \bar{e}_L}{\partial \bhl}$ 
converges to $1$ as $L \to \infty$.  This extends the relation already found
by Vespignani at al. \cite{Vespignani} for sandpiles, to the Zhang model.

\ssu{Extrapolation to $L \to \infty$.}

In this section we want to make some remarks,
based on the previous analysis, concerning the limit
of the model  when the size
$L$ tends to infinity (``thermodynamic limit'').
Note first that a mathematical definition of the thermodynamic
 limit raises serious
problems in SOC. In statistical mechanics the finite volume equilibrium 
invariant\footnote{ Note that the form
of the equilibrium measure is in general obtained from general 
principles, without reference
to the underlying microscopic dynamics, though dynamical systems
admitting the Gibbs measure as the unique invariant measure
can easily be constructed (Glauber dynamics for example).} measure is known: this is the Gibbs measure
$\frac{e^{-\beta H_{V,\bar{V}}}}{Z_V}$ where the finite volume Hamiltonian 
$H_{V,\bar{V}}$ contains interactions with the exterior
$\bar{V}=\bbbz^d\setminus V$. In SOC models
the stationary state is the result of a specific non Hamiltonian microscopic dynamics
and its explicit form is not known. Moreover, in models like the Zhang's model,
there are typically infinitely
many ergodic measures  and one has to add additional constraints
in order to select the physically relevant one. 

In statistical mechanics the infinite volume Gibbs measure
can be constructed via the Dobrushin-Landford-Ruelle (DLR)
specification  \cite{Meyer}. One attempts to find a measure $g^\ast$ on $\bbbz^d$ (or $\bbbr^d$)
 such that  the conditional measure $g^\ast_{V,\bar{V}}$, restricted to
a finite volume $V$, with specific boundary conditions in $\bar{V}$,
 is the corresponding 
finite volume Gibbs measure. Under suitable conditions 
on the interactions, one can show that this measure
exists and is unique \cite{Ruellestat}. A similar construction for SOC models can
be conceived but raises technical difficulties. One would like to construct
a measure $\hat\mu^\ast$ on $\bbbz^d$ such that the conditional measure
$\hat\mu^\ast_{\Lambda,\bar{\Lambda}}$ with zero
boundary conditions on $\bar{\Lambda}=\bbbz^d \setminus \Lambda$ is 
the stationary state of the finite volume model ($\hml$ in our case).
However, the construction of a SOC dynamical system on $\bbbz^d$ (or a Markov
process) and the computation of the corresponding
invariant measure has only been
done in specific one dimensional sandpile  models \cite{Eurandom},
and has not been yet done in larger dimension for the main SOC
models (BTW, Zhang, OFC, etc...). The main obstacle in the construction
is the presence of long range correlations which prevents the use
of classical theorems \cite{Eurandom}.

In the present paper, we will therefore not try to construct the thermodynamic
limit in a rigorous way. Rather, we will attempt to build a finite size 
thermodynamic formalism
for the Zhang's model
with the  objectives presented in the introduction.
Note that when dealing with the notion of thermodynamic limit 
one is usually faced to  the 
delicate question of the order in which the  limits 
$L \to \infty$, $T \to \infty$ are done. 
However, since usually the scaling properties
of the \textit{stationary} state w.r.t. $L$ are discussed in the SOC
literature, the limit $L \to \infty$ is implicitly
done \textit{after} the limit $T \to \infty$. This is what
we will do in this paper. Note that there are no a priori reason
why 
these two limits should commute (usually, in critical phenomena, they do not).
Note that the case $T \to \infty, L \to \infty, h=0$, there is no
dissipation at all in the limit model and consequently no stationary regime. \\

Let us now make several important remarks coming out from
the analysis made in the previous sections. For $h=0$, when
$L \to \infty$ the 
system reaches a  state where the correlation lengths $\xi_L^n$, $n=a,\tau,s$ diverge,
and where the avalanche are statistically distributed according to a power law.
In this sense, it corresponds to a critical state, which is reached spontaneously
by the only effect of the dynamics, the adiabaticity condition in the energy injection,
and the vanishing of the boundary dissipation rate. 
From the dynamics point of view,
 the positive Lyapunov exponent $\boml\log(N)$, which is also the Kolmogorov-Sinai
entropy, vanishes since the excitation rate $\boml$ tends to zero. Therefore
the asymptotic state has zero entropy. In the same time, the dissipation rate
vanishes and correspondingly, the first negative Lyapunov exponent 
tends to zero (see eq. (\ref{lambda1})). 
Consequently, the
Zhang's model loses its hyperbolic structure in the limit $L \to \infty$.
This is clearly expected since  
the loss of hyperbolicity is a necessary condition
to have an algebraic correlation decay. Indeed,  hyperbolic dynamical 
systems have exponential
correlation decay. A sufficient condition is given by the vanishing of the spectral
gap in the spectrum of the Perron-Frobenius operator (see the discussion). 
For $h > 0$, $\lambda_L(1)$ still converges to 0 but the positive Lyapunov exponent (the entropy) diverges
like $\log(N)$. On the other hand, the discussion above shows that there is
no critical state in the thermodynamic limit.
Hence,  $h$ can be used as a control parameter
allowing to tune the system to the critical regime. 
In the next sections we show that the
corresponding  phase
transition  can be handled, in the proper thermodynamic formalism, 
with the analysis tools 
of statistical mechanics (Lee-Yang zeros).
The next
section is devoted to the construction of the thermodynamic formalism
for the Zhang's model.

\su{Thermodynamic formalism.} \label{thermform}

The evolution equation (\ref{sdloc}) characterizes
 a deterministic dynamical system. One can also describe
the dynamical
evolution of the Zhang model by a Markov process with an \textit{uncountable} phase space.
However, the hyperbolicity of the finite size model, allows us
to reduce the dynamics to a Markov chain
with an most \textit{countable} phase space. 
Indeed  there are energy configurations 
equivalent from the dynamical point of view,
in the sense that their trajectories are asymptotically
indistinguishable. From the sets of equivalent configurations
one can build partitions of the phase space allowing
to encode  the dynamics symbolically by an at most
countable Markov chain. 
Moreover, in some cases,
the Markov chain is \textit{finite}. 

Furthermore, by redefining the dynamics in terms
of return maps, we construct the symbolic dynamics
in such a way that each symbol corresponds to energy
configurations undergoing \textit{the same avalanche
when a specified site is excited}. In this way,
the coding is both relevant to characterize
the microscopic evolution but also the avalanche dynamics.
The first SOC conjecture  implies that
there is a unique invariant measure in the code space, 
for the corresponding Markov chain,
corresponding to the SRB measure.  By definition,
it characterizes the microscopic dynamics. In particular
its support corresponds to the so-called ``SOC attractor''
\cite{BTW,Jensen}.
But, by construction, it also characterizes the macroscopic
avalanches distribution. This measure is a particular
example of  Gibbs measures (see appendix). We construct other examples
of Gibbs measure in subsection \ref{Gibbs}, by changing the weights
in the Markov transition matrix and discuss their connexions
with the microscopic dynamics and  the distribution of avalanche observables.

\ssu{The return maps.}\label{SystDyn}

We first redefine the  Zhang's model dynamics
in terms of return maps. 
The avalanche after the excitation of a site  maps unstable
 to  stable
configurations.
One can view this process as a mapping
from $\cM \to \cM$ where one includes the process of excitation.
 Let $T_i$
be the map $\cM \to \cM$ which associates to a stable energy configuration $\bX$
the next stable configuration resulting from an avalanche obtained by the excitation 
of the site $i$ in the configuration $\bX$. Then:

\beq
T_i(\bX) = F^{\tau(i,\bX)}(\bX+\delta \be_i), \quad \bX \in \cm
\eeq

\nid in terms of the the mapping (\ref{F}), where $\tau(i,\bX)\deq
\inf \left\{t \geq 0, F^t(\bX+\delta \be_i) \in \cm \right\}$ is the duration
of the avalanche obtained by exciting the site $i$ in the stable
configuration $\bX$.

From the mappings $T_i, i \in \Lambda$, we construct
a new (``Poincar\'e like'') dynamical system where the phase space
is $\Omega =
\Sigma^+_\Lambda \times \cM$,  where, as above, $\Sigma^+_\Lambda$ is
the set
of right infinite excitation sequences 
$\ta = \left\{ a_1,\dots, a_k, \dots \ | a_k \in \Lambda
\right\}$ but  $\bX$ is now a \textit{stable} energy configuration.
$\hX =(\ta,\bX)$ is now a point  in $\Omega$. Let $\pi_1$ and $\pi_2$ be the
canonical projection respectively on $\Sigma^+_\Lambda$ and $M$ (namely
$\pi_1(\hX)=\ta$, $\pi_2(\hX)=\bX$).

The evolution  is determined by a
dynamical system of skew product type,
 $\cT : \Omega \longrightarrow \Omega$ such that:
\beq \label{SD}
 \cT(\hat{\bX}) \deq \left( \sigma \ta, T_{a_1}(\bX) \right) \ ; \qquad
\hat{\bX}
 \deq (a,\bX)
\eeq 

Set $\hX(t)=\cT^t(\hX)$ and $\bX(t)=\pi_2(\cT^t(\hX))$.
$\bX(t)$ is now the stable energy configuration obtained after 
$t$ avalanches, starting from the energy configuration $\bX$
 when the excitation sequence is $\ta=\pi_1(\hX)$.\\

Due to the particular structure
of the mapping (\ref{F}), each map $T_{i}$, $i \in \Lambda$,  is a 
\textit{piecewise
affine map} \cite{BCK3}. 
Denote by $T_{(i,j)}$ the $j$ th affine component
\footnote{The maps $T_{(i,j)}$ play in some sense the role of the toppling operators
introduced by Dhar for Abelian sandpiles \cite{Dhar}. However, the structure
of the Zhang model is more complex since these operator do not commute
and since they do not preserve the Lebesgue measure.} of the map
$T_i$. Call $M_{(i,j)}$ the domain of definition of $T_{(i,j)}$.
There is a one to one correspondence between the set of avalanches 
and the set of affine mapping: $T_{(i,j)}$ maps
the energies configurations in $M_{(i,j)}$ to the stable
configurations resulting from the avalanche $(i,j)$.
(Recall that each
avalanche can be labeled by a double index
$\left( i
,j \right) $ where the first index refers to the site
where the avalanche starts and the second to an enumeration of the possible
avalanches starting at $i$).  
For each $i \in \Lambda$ the $M_{(i,j)}$'s, $j = 1 \dots l(i)$ constitute
a \textit{partition} of $\cm$.

 The set of points
 $\cS$ where $\cT$ is not continuous is
called the \textit{singularity set}. 
This is an union of hyperplanes 
\footnote{The number of singularity planes is finite when $L < \infty$
but it becomes infinite as $L \to \infty$ when $h=0$.}
constituting the borders $\partial\cm_{(i,j)}$
of the $M_{(i,j)}$'s. In more mathematical terms:

\beq \label{S}
\cS=\bigcup_{i \in \Lambda} \bigcup_{j = 1}^{l(i)} \partial\cm_{(i,j)}
\eeq
 
\nid where:

\beq
\partial\cm_{(i,j)}=\left\{\bX \in \cm, \ \exists k \in \Lambda,
\exists t \in \left\{1 \dots \tau(i,\bX) \right\}, F_k^t(\bX+\delta \be_i)=E_c \right\}
\eeq

The singularity set $\cS$ is therefore the set of stable energies configurations
such that some sites have an energy \textit{exactly equal to $E_c$} at some
time during some avalanche. These configurations have therefore some
site ``right
at the threshold'' at some time of their evolution. They are
therefore highly sensitive to an infinitesimal perturbation on the energy of these
sites. $\cS$ plays an important role in the dynamics, discussed in the next section
and in section \ref{FDT}.\\

It can be proved that each map $T_{(i,j)}$ is a quasi contraction.
More precisely it contracts on the subspace generated
by the the canonical basis vectors  corresponding
to the active sites, and is neutral on the remaining
part of $\bbbr^N$. Furthermore, 

\beq \label{sij}
\det(T_{(i,j)})=\epsilon^{s((i,j))}
\eeq

\nid where $s((i,j))$ is the size of the corresponding
avalanche \cite{BCK3}.

Moreover, $det(\pi_1(D\cT^t_{\hX}))=\log(N)$ since 
the shift on $\Sigma_\Lambda^+$
is conjugated to $Nx mod 1$. 
Consequently, $\pi_1(\cT)$ is expansive
with a positive Lyapunov 
exponent $\zeta_L(0)=\log(N)$.
It has also $L^d$ negative Lyapunov
exponents $\zeta_L(i), \ i=1 \dots L^d$. The  Lyapunov
exponents $\zeta_L(i)$ are related to 
the Lyapunov exponents
of section \ref{Transport} by a simple reparametrization of time.
Indeed, in the dynamical system (\ref{SD}) a unit of time corresponds to the time
elapsed between two excitations.
In the dynamical system (\ref{sdloc}) the average corresponding time is
$\frac{1}{\boml}$. Consequently,  the Lyapunov
exponents $\zeta_L(i)$ are given by:

\beq \label{zeta}
\zeta_L(i)=\frac{\lambda_L(i)}{\boml}, \ i=0 \dots N
\eeq

Since $\Omega$ has a product structure, and since
the excitation is independent of the dynamics 
on energy configurations, the invariant
measures we consider decomposes as $\hm_L^u \times \hm_L^s$ 
where  $\hm_L^u$ is the
induced measure on
the unstable direction or \textit{excitation} measure
 and $\hm_L^s$ is the induced measure on $\cM$ or measure on the 
stable energy
 configurations. In the Zhang's model
 $\hm_L^u$ is the uniform Bernoulli measure
 since the successive excited
sites are chosen independently
with a fixed rate $\frac{1}{N}$. To the SRB measure defined in section \ref{Transport}
corresponds a SRB measure for (\ref{SD}), with  product structure $\hm_L^u \times \hm_L^s$.
In the sequel 
we will keep the notation $\hml$ for the SRB measure of (\ref{SD}).

\ssu{Symbolic dynamics.}\label{Symbolic}

In this section we  build a symbolic coding of the dynamics,
relevant to characterizes both  the microscopic dynamics and
the avalanches dynamics. For, one has first to build
a suitable partition $\cP$ of the phase space $\Omega
=\Sigma^+_\Lambda \times \cM$.

A \textit{rooted avalanche} is a pair $(\cC,B)$ where $\cC$ is an avalanche
and $B \subset \Omega$ such that for all points $\hX \in B$ the only possible
avalanche is $\cC$. Equivalently, if $\cC =(i,j)$ then $\pi_2(B) \subset M_{(i,j)}$ and
$\pi_1(B) = [i]$,
where $[i]$ denotes the set of excitation sequences whose first
digit is $i$ (cylinder set) $\left\{ \ta \in \Sigma^+_\Lambda \ | \ 
a_1=i\right\}$.
A system $\left(\cC_\omega,B_\omega\right)_{\omega \in \cI}$ of rooted avalanches,
where the set
of indexes $\cI$ is possibly countable,
is called \textit{complete} if $\bigcup_{\omega \in \cI}B_\omega = \Omega$ and
$B_\omega \cap B_ \omega'= \emptyset$ for $\omega \neq \omega'$. 
Hence, $\left\{B_ \omega\right\}_{\omega \in \cI}$
is a  partition of $\Omega$. By construction
to each element $\omega$ of this partition 
 corresponds a unique 
avalanche $(i,j)$. Equivalently, to
each $\omega$ one can associate 
a
double
index $\left( i\left( \omega \right)
,j\left( \omega \right) \right)$ where the first index 
refers to the site
where the avalanche starts and the second to
the corresponding avalanche. 
Since $\pi_2(B) \subset M_{(i,j)}$ 
the avalanche $\left( i,j\right) $ corresponds
in general to several $\omega$'s. However, the partition 
$\cP_{nat}= \left\{[i] \times M_{(i,j)}\right\}_{i \in \Lambda, j=1 \dots l(i)}$,
called the \textit{natural} partition in the sequel,
is such that
there is a \textit{one to one} correspondence
between the symbols $\omega$ and the avalanches. By definition
any complete system of rooted avalanche corresponds to
a partition generated from $\cP_{nat}$. 

To each  complete system of
rooted avalanches $\left(\cC_\omega,B_\omega\right)_{\omega \in \cI}$, 
we associate a transition
matrix $\cA=(\cA_{ij})_{i,j \in \cI}$ such that $\cA_{ij}=1$
if $B_j \cap \cT^{-1}(B_i) \neq \emptyset$ (the transition
$i \rightarrow j$ is legal) and $\cA_{ij}=0$
otherwise, and a transition graph $\cG$
with  vertices $\omega \in \cI$ and oriented edges $i \rightarrow j$
for all pair $(i,j)$ such that $\cA_{ij}=1$.
This provides a \textit{symbolic dynamics} description of the Zhang's
model where the trajectory
of a point is represented by a legal sequence
of symbols $\dots \omega_1\omega_2 \dots \omega_n \dots$
corresponding to the partition elements that this point meets
in its history. However, except for special partitions called 
generating partitions, the coding 
is in general not unique.  

The 
transition graph $\cG_{nat}$ of the natural partition characterizes the legal compositions 
between avalanches (resp. mappings).
However, though the natural partition is the most suitable
to describe the avalanches evolution it does not have,
in general, the 
\textit{Markov partition property} 
allowing to encode the dynamics in a \textit{Markov}
chain. We say that a complete system of rooted avalanches
(resp. the corresponding  partition
$\left\{B_ \omega\right\}_{\omega \in \cI}$) has the  \textit{Markov
partition property} if, for all couple $(i,j)$,
such that $\cA_{ij}=1$ :

\bea\label{Markov}
(i) \quad \pi_1(\cT(B_i)) &\supset& \pi_1(B_j)\\
(ii) \quad \pi_2(\cT(B_i)) &\subset& \pi_2(B_j)
\eea

By construction the property (i) always holds.   
If (ii) holds then  the corresponding 
complete  system of rooted avalanches has the following properties:

\bit
\item(i) Every (legal) finite path $ \omega_1, \dots, \omega_n$, $\omega_k \in \cI$ 
corresponds to a \textit{legal
sequence of avalanches} $(i(\omega_1),j(\omega_1)), \dots, (i(\omega_n),j(\omega_n))$.

\item(ii) For any $\hX$ each orbit segment 
$\left\{\cT^l(\hX) \right\}_{1 \leq l \leq n}$ is realized as a path
of length $n$ on the Markov transition graph.
\eit

Except for a non generic set, 
the trajectory of any point $\hX$ in the phase space $\Omega$
is \textit{uniquely} encoded by  an infinite sequence of symbols
corresponding to the partition elements visited by $\hX$. 

The physical interpretation of the property (ii)
is that,
starting from any energy configuration
in one element of the partition and exciting a given site
the next avalanche is \textit{uniquely} determined by the next excited
site. Consequently, the natural partition $\cP_{nat}$ does not obey  (ii))
in general, though, in some specific examples, there exists
a finite iterate of $\cT$, $\cT^t$ such that $\cT^t(\cP_{nat})$
has the Markov property. This is the case of the one dimensional
Zhang model where $E_c \in ]1,2],h=0,\epsilon=0$ \cite{BCK3}.
In the general case, in order to have a coding
relevant both for the avalanche dynamics and having
the Markov property, the strategy is to build a partition from
the natural one, by cutting it in sufficiently small sub pieces.
However the resulting partition can be countable. \\

The condition (ii) is  rather strong since it deals with all points
in $\Omega$. Since we essentially want
to have a symbolic description of the \textit{asymptotic} dynamics,
namely on the attractor which is a subset in $\Omega$,
it is in fact enough to prove the existence of a finite Markov
partition in the usual sense \cite{Katok}.
 This is insured by the existence of a \textit{local stable
manifold} of sufficiently large diameter, for almost every
point in $\Omega$ \cite{BCK3}. Recall that the local stable manifold $\cW^s_{loc} (\hX)$
 of $\hX \in \Omega$  is the largest connected set
such that $\forall \hY \in \cW^s_{loc}, d(\cT^t(\hY),\cT^t(\hX)) \to 0$
when $t \to \infty$, $d$ being some distance on $\Omega$. Therefore,
the trajectory of any point belonging to  $\cW^s_{loc}(\hX)$
is asymptotically identical to the trajectory of $\hX$ and these
points are equivalent from the dynamical point of view. Equivalently,
if $\hX$ has a local stable manifold of diameter larger than some $ \eta >0$ then
a small perturbation of size $\leq \eta$ will be asymptotically damped
(see section \ref{FDT} for a physical interpretation of this effect.).

Were the map $\cT$ be  regular, were the existence of local stable
manifolds and Markov partition be  insured by the standard
theorems on
regular ($\cC^{1+\alpha}$) hyperbolic dynamical system \cite{Pollicott}.
However, $\cT$  is not continuous on the singularity set $\cS$ and some 
points  have no local stable manifolds.
The main problem is to estimate the $\hml$ measure of the ``bad'' points
having no local stable manifold. The following result is useful.

\bp Let $\cU_\delta(\cS) =  \left\{\hY \in \Omega \ | \ d(\hY,\cS) \leq \delta  \right\}$
be the $\delta$-neighborhood of $\cS$. Assume that for $\delta>0$ sufficiently small:

\beq \label{majormu}
\hml\left[\cU_\delta(\cS) \right] \leq C \delta^\alpha
\eeq

\nid some $C>0$, some $\alpha>0$. Then $\hml$ almost every point 
has a local stable manifold of positive diameter.
\ep

\bpr
The existence of a local stable manifold of positive diameter for a point
$\hX$ in the support of $\hml$ is ensured if one can find a number $0 >  \gamma > \lambda_L(1)$,
and a time $t_0 <  \infty$ such that $\forall t > t_0$,
$d(\cT^t\hX,\cS) > e^{\gamma t}$. Indeed, in this case a sufficiently small ball around
$\hX$ is asymptotically contracted faster than it approach the singularity set,
and consequently, all the points of this ball belong to the stable manifold
of $\hX$. Consequently, the set of bad points is included in the
set :
$$
\left\{  \hX \ | \ \forall t_0 \geq 0, \ \exists t \geq t_0, \ d(\cT^t\hX,\cS) \leq e^{\gamma t} \right\}
 = \bigcap_{t_0=0}^\infty \bigcup_{t=t_0}^\infty
\cT^{-t}\left( \cU_{e^{\gamma t}}(\cS)\right)
$$
\nid for  some $\gamma$ such that $0 >  \gamma > \lambda_L(1)$.
From the Borel-Cantelli lemma the measure of this set is zero
provided the series
 $\sum_{t=0}^\infty \hml\left[ \cU_{e^{\gamma t}}(\cS)\right]$
converges. This is true if the condition (\ref{majormu}) holds.
\epr\\

We have not established yet a mathematical proof of the condition  (\ref{majormu})
but we have strong numerical evidences presented in the section \ref{FDT}. 
In the sequel we will therefore assume that $\hml$ almost every point
as a stable manifold of positive diameter. Then, it is possible to generate
a finite Markov partition $\cP$ encoding the asymptotic dynamics
and such that each symbol corresponds to a unique avalanche.
Note however that \textit{several symbols}
may  correspond to the \textit{same avalanche}. Nevertheless, though an avalanche $\left( i,j\right) $ can
be represented by several  partition-elements, 
in order to simplify the notation we will however
 identify $\omega$ and the avalanche
$\left(i(\omega),j(\omega)\right)$ whenever it makes
no confusion.
Denote by $\tom=\omega_{-n} \dots \omega_{-1} \omega_0
\omega_1\omega_2 \dots \omega_n \dots$, where
$\cA_{\omega_n,\omega_{n+1}}=1, n \in \bbbz$, a legal bi-infinite
sequence. Call $\cX$ (resp. $\cX^+$) the set of 
bi-infinite (resp. right infinite) legal sequences.
There is a one-to-one correspondence (up to a 
set of zero Lebesgue measure), denoted by $\simeq$,  between a 
symbolic sequence $\tom$ and a point $\hX$ in the phase space.
Let $\pi =(\pi ^{+},\pi ^{-})$ be the corresponding conjugating mapping, such that
if $\tom \simeq \hX=(\ta,\bX)$ then $\pi ^{+}(\tom)=\ta$
and $\pi^{-}(\tom)=\bX$. Note that $\pi ^{+}$ projects on the
unstable direction ($\Sigma_\Lambda^+$),
 while $\pi^{-}$ projects on the stable space ($\cm$).
Write $[\tom]_n$ for a n-cylinder (this is
the subset of $\cX$ where the sequences have
the same $n$ first digits as $\tom$). 
Denote by $\sigma_\cA$ the shift on $\cX$. 
The forward orbit of $\tom$ under $\sigma_\cA$
encodes the excitation sequence by  definition and
the backward orbit of $\tom$ encodes the point in the energy configuration
space ${\cal M}$.

\ssu{Markov chain and SOC state.} \label{Statstat}

We construct a Markov chain by defining a transition kernel
$\cW$ from the matrix $\cA$. $\cA$ has the following property.
 Let $D_{\omega }^{+}$ be the set of follower
elements of $\omega $ (and $D_{\omega }^{-}$ the set of predecessors of $%
\omega $). $\#D^+_\omega \geq N, \ \forall \omega$ since after
the avalanche corresponding to $\omega$ one can excite any site
in $\Lambda$. Moreover, the Markov partition property
(\ref{Markov} ii) implies that if $\beta ,\gamma $ $\in D_{\omega }^{+}$ and 
$\beta \neq \gamma$, then $i(\beta) \neq i(\gamma)$.
 Hence, $\#D_\omega=N, \ \forall \omega$. Consequently,
$\cA$ has \textit{exactly $N$ non zero component per row}, corresponding
to each of the $N$ possible choices of an excited site.

Since the excited site of $\Lambda $ is chosen with respect to the
uniform Bernoulli measure  the transition probability  of the Markov-Partition
for any edge from $\omega $ to $D_{\omega }^{+}$ 
is constant, equal to $\frac{1}{N}$. Therefore the transition kernel
is :

\beq \label{W}
\cW = \frac{1}{N}\cA
\eeq

As a consequence,
$\cW$ is a \textit{sparse} matrix. Indeed, the number
of symbols $\omega$ is at most equal to the number of possible avalanches,
which is bounded from below by the largest avalanche size (resp.
duration, area) $\xi_L^s$. Since, for $h=0$, $\xi_L^s \sim L^{\beta_s}, \ \beta_s > d$, 
 the proportion
of non zero entry on each row,
$r_L=\frac{N}{\#\cI}  L^{d-\beta_s}$, tends to zero as $L \to \infty$.
This remark justifies somehow to approximate the transition matrix
by a \textit{random} matrix. This could be used as an approximation
to determine the spectral gap of $\cW$ (see the discussion).\\

The set of symbols $\cI$ decomposes into transient and recurrent
nodes. Call $\cR(\cI)$
the  set of recurrent nodes and
 $\cN_L  \deq \#\cR(\cI)$ the number of elements
in $\cR(\cI)$. The first SOC  conjecture
requires that the set of recurrent node is
irreducible.  This is \textit{not} in contradiction
with the sparseness of $\cW$. For example, it can be proved
that for sparse random matrices, with $K(m)$ uniformly distributed
non zero entry per row where $m$ is the size of the matrix,
the set of nodes is almost-surely constituted
by \textit{one} irreducible recurrent cluster as $m$ tends to infinity
provided $K(m)>\log(m)$ \cite{Flajolet}.

The equivalent of the convergence property (\ref{SRBmix})
is :

\beq \label{Mixing}
\ml=\lim_{t \to \infty} \tilde{v}\cW^t
\eeq

\nid where  $v$ is any initial probability distribution on $\cI$.
This property  holds when $\cW$ is mixing  \footnote{There exists
$n \geq 1$ s.t. for each pair $i,j$ $\cW^n(i,j) >0$, i.e. there is a path
connecting each node $i,j$.}. 
In this case $\cW$ has a unique eigenvalue $1$
and a unique left eigenvector $\ml$, corresponding to the
invariant probability distribution of the Markov chain \cite{Seneta}. 
$\ml(\omega)$ is non zero only if $\omega \in \cR(\cI)$.
Furthermore, there is a spectral gap between the second eigenvalue with
the largest modulus and $1$. The gap gives the exponential correlation
decay and also the rate of convergence to equilibrium. It is
expected that the gap vanishes as $L \to \infty$ for $h=0$
and stays positive for $h>0$.

The invariant probability distribution
$\ml$ characterizes the probability of occurrence
of any recurrent symbol at stationarity. This is therefore
the fundamental object characterizing the SOC state.
In particular the probability distribution of 
an avalanche observable is given by :

\beq \label{PL}
P_L(n,h) = \sum_{\omega \in \cI, n(\omega)=n} \ml(\omega)
\eeq

\nid  Recall that $n(\omega)$ stands for $n(i(\omega),j(\omega))$. This is
the value that the observable $n$
takes in the avalanche $i(\omega),j(\omega)$. 
 Note that $\ml, \hml$
depend on $h$ but we dropped this dependence in order
to simplify the notations. The moments of
$n$ are given by:

\beq\label{nLp}
\left<n^q\right>_{L,h} = \sum_{\omega \in \cI} 
n^q(\omega)\ml(\omega)
= \sum_{n=0}^{\xi_L^n} n^q
\sum_{\omega \in \cI, n(\omega)=n}\ml(\omega)
=\sum_{n=0}^{\xi_L^n} P_L(n,h)n^q
\eeq

More generally, the joint probability $\tml$ on the space of infinite symbolic
sequences  $\cX$ is obtained from the Chapman-Kolmogorov
equation. For any cylinder $[\tom]_T=\omega_1 \dots \omega_T$:

\beq \label{jointProb}
\tml([\tom]_T)=
\ml(\omega_1)\cW_{\omega_1\omega_2}\cW_{\omega_2\omega_3}\dots\cW_{\omega_{T-1}\omega_T}
\eeq

Note that $\tml$ is the measure of maximal entropy $\log(N)$.

The measure $\tml$ projects down to the SRB measure 
$\hm_L$ on $\Omega$ \ via $\hm_L =\tml\circ \pi ^{-1}$
with marginals $\hm_L^u=\tml\circ \pi ^{+}$ and $\hm_L^s=\tml
\circ \pi ^{-}$.  
For the following we will therefore make no distinction between 
the average with respect to $\tml$ or with respect to $\hm_L$. 
From the measure $\tml$ one can 
compute all the time correlations 
(therefore the joint probability on the space
of trajectories of $n(t)$)
 of the observable $n$ (see eq. (\ref{Plconj})).\\

The knowledge of the transition matrix $\cW$ allows us to determine
the invariant measure $\ml$ and the spectral gap. However its structure
is not known in general. An interesting, but very specific case, occurs
when the number of predecessors of $\omega$, $\#D_\omega^-$,
is also equal to $N$, $\forall \omega$. This happens for
example in the one dimensional Zhang model where $E_c \in ]1,2],
\epsilon=0,h=0$ \cite{BCK3}. Then $\cW$ is a bi stochastic
matrix and  $\ml$ is the \textit{uniform measure}, $\ml(\omega) =\frac{1}{\cN_L}$, on $\cR(\cI)$.
However, in general, $\#D_\omega^-$ depends on $\omega$.

\ssu{Thermodynamic formalism, Gibbs measures
and generating function of avalanche size distribution.} \label{Gibbs}

In this section we  use the thermodynamic
formalism (see appendix) to construct Gibbs
measure, by choosing different families
of potentials. These measures are relevant to characterize microscopic properties
(like the fractal structure of the SOC attractor), but
also the  avalanche distributions.

The SRB measure $\hml$ is the equilibrium state
which maximizes the potential $-\log(\det(\pi^+(D\cT_{\hX})))$
\cite{Keller}.
In the Zhang model this potential is a constant $-\log(N)$.
It follows that the SRB measure maximizes the entropy.
 Since the maximal
metric entropy is the topological entropy $\log(N)$, it also follows that the
SRB measure has zero pressure. 

Due to the product structure of the Zhang's model:

 \beq \label{det}
log\left(det(D\cT_{\tom})\right)=\log(N)+\log\left(det(DT_{\omega_1})\right)
\eeq

\nid where $DT_{\omega_1}$ stands for 
$DT_{\left(i(\omega_1),j(\omega_1)\right)}$.
Let us introduce a first family of potentials:
 
\beq\label{phi1}
\phi^1_\beta(\tom)=-\log(N)+\beta\log\left(det(DT_{\omega_1})\right)
\eeq

\nid and denote the corresponding family of equilibrium states 
by $\mu_\beta$ in the sequel.
The potential $\phi^1_\beta$ are Bernoulli potentials (they depends only on the first symbol 
in $\tom$). Consequently, they trivially match the condition (\ref{decay}) in the appendix.

The case $\beta=0$ corresponds to the SRB measure.
The quantity 
$\log(det(DT_{\omega_1}))$
is the volume contraction in the space of energy configuration, induced
by the relaxation dynamics. It is related to the energy transport in the lattice
and to the avalanche size $s(\omega_1)$.
Indeed,  eq.(\ref{sij})  implies:

\beq 
\phi^1_\beta(\tom)=-\log(N)+\beta \log(\epsilon)s(\omega_1)
\eeq

The corresponding  partition function (\ref{ZT}) writes:

\beq \label{ZLmarg}
Z^1_T(\beta) = N^{-T} \sum_{\omega \in \chi^+_T} 
\epsilon^{\beta\sum_{t=1}^T
s(\omega_t)}
\eeq

The corresponding topological pressure is:

\beq \label{F1}
\cF^1_L(\beta) \deq \lim_{T \to \infty} \frac{1}{T}\log\left(Z^1_T(\beta)\right)
\eeq

Then from eq. (\ref{aimantation}),

\beq \label{sL}
\left. \frac{\partial \cF^1_L(\beta)}
{\partial \beta}
\right|_{\beta=0}= 
 \log(\epsilon)\left<s\right>_{L,h}
\eeq

This equation is analogous to the one giving the 
magnetization when deriving the free energy w.r.t.
to a uniform local field. 
It relates the average
volume contraction rate given by the sum of negative Lyapunov exponents
$\zeta_L(i), \ i=1 \dots L^d$,
to the average avalanche size \cite{BCK3}

\beq \label{contav}
\sum_{i=1}^N \zeta_L(i)=\log(\epsilon)\left<s\right>_{L,h}
\eeq

The second derivative of the free energy (\ref{F1})
with respect to $\beta$ is:

\beq  \label{d2F1}
\left. \frac{\partial^2 \cF^1_L(\beta)}
{\partial \beta^2}
\right|_{\beta=0}= 
 \lim_{T \to \infty} \frac{1}{T} \sum_{t=1}^T \left<s_{\omega_t}s_{\omega_0}\right>_{L,h}
-\left<s\right>^2_{L,h}
\eeq

This equation gives therefore the variance, along
a typical trajectory, of the Gaussian
fluctuations of the avalanche size (resp. contraction rate) 
around the average values. In section \ref{LeeYang} we discuss the failure of
differentiability of $\cF_L(\beta)$ for $h=0$, in the thermodynamic limit,
which implies that (\ref{d2F1}) diverges and that the central limit theorem
is violated.\\

More detailed informations about the probability distribution of avalanches
size, or more generally of any other avalanche observable $n$,
can be obtained by introducing in the potential the formal equivalent of
a local magnetic field $\eta$ which allows, in statistical mechanics, 
to compute the local magnetization and higher order cumulants.
A natural extension of (\ref{phi1}) is therefore the time dependent potential:

\beq \label{phi1ext}
\phi_\eta(t,\tom)=-\log(N)+\eta_{t}n({\omega_t})
\eeq

\nid where $\eta=(\eta_1, \dots, \eta_t, \dots)$, and
the partition function :

\beq \label{ZLconj}
Z_T(\eta_1,\dots,\eta_T) =  N^{-T} \sum_{\omega \in \chi^+_T} 
e^{\sum_{t=1}^T \eta_{t}n({\omega_t})}
\eeq

This expression reduces to (\ref{ZLmarg}) when $\eta_t=\beta\log(\epsilon)$, $n=s$.
Set :

\beq\label{Plconj}
\cF_L(\eta)=\lim_{T\to\infty} \frac{1}{T}\log(Z_T(\eta_1,\dots,\eta_T))
\eeq

$\cF_L(\eta)$ generates
the $k$-time correlations of the observable $n$:

\beq
\left. \frac{\partial^k \cF_L(\eta)}
{\partial \eta_1\dots \partial \eta_k}
\right|_{\eta=0}= \left<n_1;n_2;\dots;n_k\right>_{L,h}
\eeq

 In particular, the cumulants $C_L(k)$ of 
the marginal distribution $P_L(n)$
are given by:

\beq\label{PlMarg}
\left. \frac{\partial^k \cF_L(\eta)}{\partial \eta_1^k}
\right|_{\eta_1=0}= C_L(k)=\left. \frac{\partial^k G_L(t)}
{\partial t^k}
\right|_{t=0}
\eeq

\nid where :

\beq \label{GL}
G_L(t)\deq \log(\left<e^{t n} \right>_{L,h}) = \log(\sum_{n=0}^{\xi_L^n(h)}P_L(n,h)e^{tn})
\eeq

\nid is the generating
function of the cumulants of $P_L(n,h)$. 

$\cF_L(\eta)$
plays a similar  role as the generating functional in quantum field theory
and $\eta$ acts as a conjugated field to the observable $n$. 
It
also corresponds
to a time dependent perturbation of the trajectories or,
equivalently, to a time dependent potential.
It must in particular be emphasized that the
topological pressure in (\ref{Plconj}) contains 
\textit{all informations} about the scaling property
of the probability distribution on the space
of trajectories of the avalanche observable $n$.
 Consequently, a scaling theory, analogous to
 statistical mechanics seems achievable.
 Formally, when $h=0$, one can associate to each local field $\eta_i$ an exponent
$y_i$ and seek a scaling form for $\cF_L(\eta)$ such as :

\beq \label{MSR}
\cF_L(\eta)=L^\beta \cH(L^{y_1}\eta_1, \dots,L^{y_n}\eta_n, \dots)
\eeq

\nid where $\cH$ is a universal function.
This provides the scaling of the $n$ points cumulants.
However, one has introduced an infinite number of fields
and this formula raises the question: what are the relevant
fields (in the renormalization group sense, that is
unstable directions for the renormalization flow)
or in other words which scaling exponents
 define the universality classes in SOC ? In classical
critical phenomena the renormalization group approach, and
the renormalization properties of the Hamiltonian and free
energy \cite{Fisher} allow to select the relevant fields
and two scaling exponents are extracted; $\alpha$, the specific heat
exponent and $\eta$ (correlation length).  In the SOC case 
the main difficulty is  to adapt the renormalization schemes and
to select the relevant fields in order to build the scaling
theory.  One possible strategy
could example be to 
adapt the Dynamically Driven
Renormalization Group (DDRG) procedure developed by Pietronero et al.
\cite{Pietronero}.
We believe
that  exponents analogous to $\alpha,\eta$ can be extracted for the Zhang model
from the scaling  of the expectation with respect
to $\tml$ and of the spectral gap of $\cW$.

For the generating function (\ref{GL}) of the marginal probability distribution $P_L(n)$ 
the scaling form (\ref{MSR}) reduces to :

\beq\label{Kada}
G_L(t)=L^{\beta_n(1-\tau_n)}G(tL^{\beta_n})
\eeq

\nid where $G$ is a universal function. This form corresponds
to the Kadanoff et al. finite-size scaling ansatz \cite{Kadanoff}.
In particular, the cumulant $C_L(q)$ has a scaling factor :

\beq
\sigma(q) \deq \lim_{L \to \infty} \frac{\log(C_L(q))}{\log(L)}
\eeq

\nid which is a linear function\footnote{ 
Recently, it has been argued that the finite-size form (\ref{Kada}) might be violated
in several models like the BTW model \cite{Tebaldi} or the Zhang's model
\cite{Vespignani2}.
 Alternative scaling 
where $\sigma(q)$ is a non linear function of $q$ have been proposed 
\cite{Tebaldi}.
The conclusions in \cite{Tebaldi,Vespignani2} 
were however essentially supported by numerical simulations and theoretical 
results are still missing. In particular the numerical bias
induced by numerics were not discussed \cite{CM}.
 Such a scaling entails very singular properties
for the asymptotic distribution function (resp. the asymptotic topological pressure, if
it exists) and, consequently, for the asymptotic
dynamical system. This opens interesting questions and perspective discussed 
in a forthcoming paper.} of $q$, $\sigma(q)=\beta_n(q+1-\tau_n)$, for 
$q\geq \tau_n-1$. 

Note that the measured exponents $\tau_n \in ]1,2[$ for
the usual avalanche observables. Consequently,
$\sigma(q)>0$ for $q\geq \tau_n-1$, and the moments
of order $q\geq \tau_n-1$ diverge in the limit $L \to \infty$. This implies
\textit{a loss of analyticity} of the limiting generating function.
As discussed below and in \cite{CM} this effect is manifested by
 a Lee-Yang phenomenon. In particular, the properties of the critical
zeros are directly related to the $\sigma(q)$'s. In the case (\ref{Kada})
the distribution of zeros is characterized by the exponents $\beta_n,\tau_n$,
but  relations linking the Lee-Yang zeros distribution to the
$\sigma(q)$'s can be derived for more general scaling \cite{CM}.\\

Let us now discuss another extension
of the potential (\ref{phi1}).
At
a microscopic level,
this potential characterizes the local contraction along a trajectory.
Since the local contraction is related to the fractal structure
of the support of the invariant measure, this remark suggests that it should
possible to make a connexions between the multifractal spectrum
of the measure $\tml$ and the avalanches  distribution.
More precisely, the dynamical system (\ref{SD}) may be considered as
 a probabilistic graph
iterated function system (see \cite{Falconer} for a review)
with maps $\left\{T_\omega \right\}_{\omega \in \cI}$,
with transition graph $\cA$ and where the probability 
 is the SRB measure \cite{BCK3}. Indeed, though the maps
are only quasi contractions any finite composition of those maps
along the graph is a contraction. In order to apply the theory developed
for iterated function systems  one needs
additional conditions (open set condition, or strong separation
condition) which are discussed in \cite{BCK3}.
Were the Zhang's model a  conformal iterated functions systems 
would  the potential which allows to compute the
multifractal spectrum be  given by:

\beq \label{phi2}
\phi^2_{\beta,q}(\tom)=-q\log(N)+\beta \log(\epsilon)s(\omega_1)
\eeq

The multifractal spectrum is the Legendre transform 
of the function $D(q)=\frac{\beta(q)}{q-1}$  
where $\beta$ is such that\cite{Falconer}:

\beq
\cF^2_L(\phi_{\beta(q),q})=0
\eeq

Though the Zhang's model is not conformal
since the contraction is not uniform in the phase
space (see the Lyapunov spectrum
Fig. \ref{graphLyap}), the potential (\ref{phi2})
might be useful when considering the marginal
energy distribution of one site. This is under
current investigations and will be published elsewhere
 \cite{Sandra}.

To compute the multifractal spectrum of
$\tml$  a more elaborate potential is required
\cite{Falconer2}. In particular, the corresponding thermodynamic formalism
is not additive, like in the appendix, but sub-additive.
 The framework for such
a generalized thermodynamic formalism is described in \cite{Falconer3}
and in greater generality in \cite{Barreira}.
 Let  $\alpha_i(\tom,k), \ i=1 \dots
N, k=1 \dots \infty$ be the singular values of the map 
$T_{\omega_k} \circ \dots \circ T_{\omega_1}$ ordered
such that $1 \geq \alpha_1(\tom,k) \geq \dots \geq \alpha_N(\tom,k) >0$
and remark that there is a finite $k$ such that, whatever $\tom$, $\forall
l > k$, 
$1 > \alpha_1(\tom,l)$. Define  :

\bea \label{phinonconf}
g_{\beta}(\tom,k)&=&\alpha_1(\tom,k)\alpha_2(\tom,k)\alpha_{j-1}(\tom,k) 
\left(\alpha_{j}(\tom,k)\right)^{\beta-j+1}, \quad if \ 0<\beta\leq N \label{phinonconf1}\\
g_{\beta}(\tom,k)&=&(det(T_{\omega_k} \circ \dots \circ
T_{\omega_1})^{\frac{\beta}{N}})=
\epsilon^{\frac{\beta}{N} \sum_{l=1}^k s(\omega_l)}, \quad if \ \beta\geq N \label{phinonconf12}
\eea

\nid where, in eq. (\ref{phinonconf1}), 
$j$ is the integer such that $j-1<\beta \leq j$. Consider the the (sub-additive) potential:

\beq \label{phi3}
\phi^3_{\beta,q,k}(\tom)=-q\log(N)+\log(g_{\beta}(\tom,k))
\eeq

\nid and define the pressure by;

\beq
\cF^3_L(\phi_{\beta,q})=\lim_{k \to \infty} \frac{1}{k} log \sum_{\omega \in \chi^+_k}
\phi^3_{\beta,q,k}(\tom)
\eeq

\nid Then there exists a unique $\beta\equiv\beta(q)$ such that 
the topological pressure $\cF^3_L$ 
 vanishes \cite{Falconer2}. The numbers $D(q)=\frac{\beta(q)}{q-1}$
are the Reyni dimensions, and the multifractal spectrum is
the Legendre transform of $D(q)$.
We are currently investigating the  relations between this potential
and the corresponding multifractal properties to
macroscopic transport properties \cite{Sandra}.   
In particular, we are trying to interpret the singular values
in terms of avalanches properties.

\su{Some remarks about the thermodynamic
limit.} \label{ThLim}

In this section we want to discuss
several effects observed when the size
$L$ tends to infinity (``thermodynamic limit'').
 These effects can be anticipated
with the
 tools and concepts developed in the previous sections,
though they do not allow yet to handle them
analytically. Consequently, most of the results
presented are conjectured from the thermodynamic
formalism description and are numerically checked.

\ssu{Lee-Yang phenomenon.}\label{LeeYang}

Rather than attempting to define the thermodynamic limit for the Gibbs
state one can, as usual in equilibrium statistical mechanics of phase transitions,
focus on the analyticity properties of the thermodynamic potential
(free energy). More precisely,  from eq. (\ref{Plconj}),
 the topological pressure is nothing but the generating function for
the joint distribution of the avalanche observable $n$. This quantity
certainly exists for finite $L$ and has to be also defined in the thermodynamic
limit if we admit that there exists a
limit probability  when $L \to \infty$.
However, if a critical state is indeed achieved, the finite size topological pressure
 should therefore develop singularities as $L\to \infty, h=0$. 

 The topological
pressure (\ref{Plconj}) is a complicated object to handle, even numerically.
However, one can argue that if the generating function (\ref{PlMarg}) of the \textit{marginal}
distribution develop singularities in the thermodynamic limit, then $\cF$
develops singularities as well. In equilibrium statistical mechanics,
a standard way to handle the singularities of the free energy and the scaling
as $L \to \infty$ is the study of the Lee-Yang zeros \cite{LY}.
In many examples the partition  function of the
 finite size systems  is a polynomial in a variable $z$ 
which typically depends on control parameters like the temperature 
or the external field. 
Since all coefficients are positive there is no zero
 on the positive real axis. However, in the thermodynamic limit, 
at the critical point, some zeros pinch the real axis at $z=1$, 
leading to a singularity in the free energy. 
The finite-size scaling properties of the leading zeros and of the density 
of zeros near $z=1$ determine the order of the transition \cite{Janke} 
and also the critical exponents in the case of a second order phase transition 
\cite{Kim}.
In this paper, we show numerically that the same effect arises for the generating
function of $P_L(s,h)$ for $h=0$, while there is no Lee-Yang phenomenon
for $h>0$. This property is not specific to the observable $s$
or to the Zhang's model. We showed indeed in \cite{CM} that this property arises
as soon as a probability distribution $P_L(n), \ n =1 \dots \xi_L^n < \infty$
 converges to a power law
$\frac{K}{n^\tau}$, $n=1 \dots \infty$ when $L \to \infty$.
 Furthermore, when  $1<\tau<2$ (which is the case
for the usual avalanche observables and all
the models), interesting anomalous finite size scaling effects
are observed.  \\

 Since
$\xi_L^n(h)$
 is finite for finite $L$, the generating function
 
\beq
\cZ_L(z)=\sum_{n=0}^{\xi_L^n(h)} z^nP_L(n,h)
\eeq
 
\nid where $z \in \bbbc$, is a polynomial of degree $\xi_L^n(h)$.
In particular, since $\cZ_L(z)$ is an analytic function
of $z$ in the complex plane, all its moments exist. 
Since all $P_L(n,h)$  are positive $\cZ_L(z)$ has no zero
on the positive real axis for finite $L$. Consequently the log-generating
function $log(\cZ_L(z))$ is well defined on  $\bbbr^*_+$. More precisely,
$\cZ_L(z)>0$ for $z$ in a small neighborhood of $\bbbr^*_+$ where $G_L(t) = log(\cZ_L(e^t))$
is an analytic function of $z$ (resp. $t$).

For finite $L$, $\cZ_L(z)$ has $\xi_L^n(h)+1$ zeros in $\bbbc$,
that are either real $\leq 0$, or complex conjugate.
Denote them
by $z_L(k), k=0 \dots \xi_L^n$ and order them such that
$0 < |z_L(1)-1| \leq \dots \leq |z_L(k)-1| \leq \dots \leq |z_L(\xi_L^n+1)-1|$.
Write $z_L(k)=R_L(k)e^{i\theta_L(k)}
=e^{t_L(k)}$.\\

We report now the following numerical observations
for the avalanche size distribution. A general theory and analytical results, extending
to other models of SOC, are developed in 
\cite{CM}.

\bit

\item The zeros
lie on a curve $\cC_L$ in the complex plane
which  accumulate to the unit  circle.
This curve is however \textit{not a circle}. In particular, it has
a ``cusp like shape'' in the neighborhood of $z=1$
(Fig. \ref{patternzeros}a). An analytic
form is given in \cite{CM}.

\bef
\bc
\begin{minipage}{6cm}
\epsfxsize=6cm
\epsfysize=6cm
\epsffile{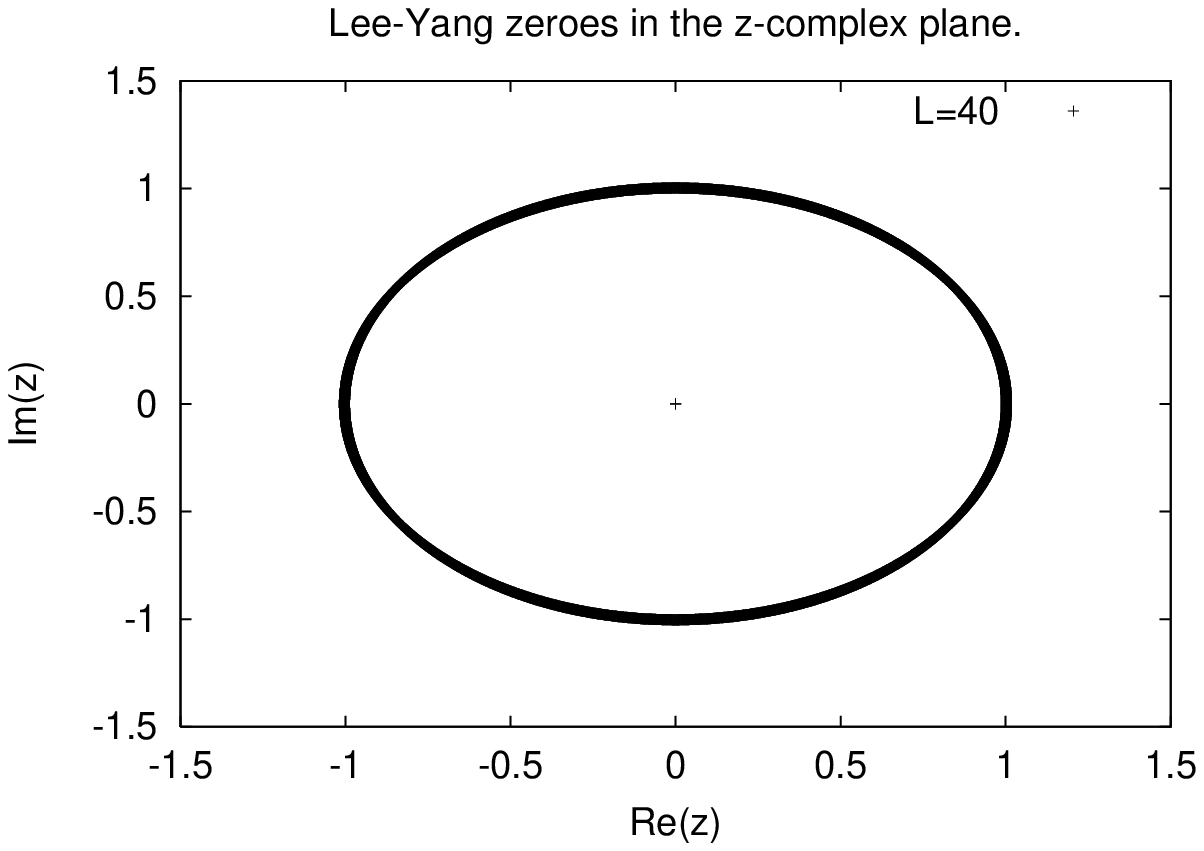}
\end{minipage} \hspace{1cm}
\vspace{0.3cm}
\begin{minipage}{6cm}
\epsfxsize=6cm
\epsfysize=6cm
\epsffile{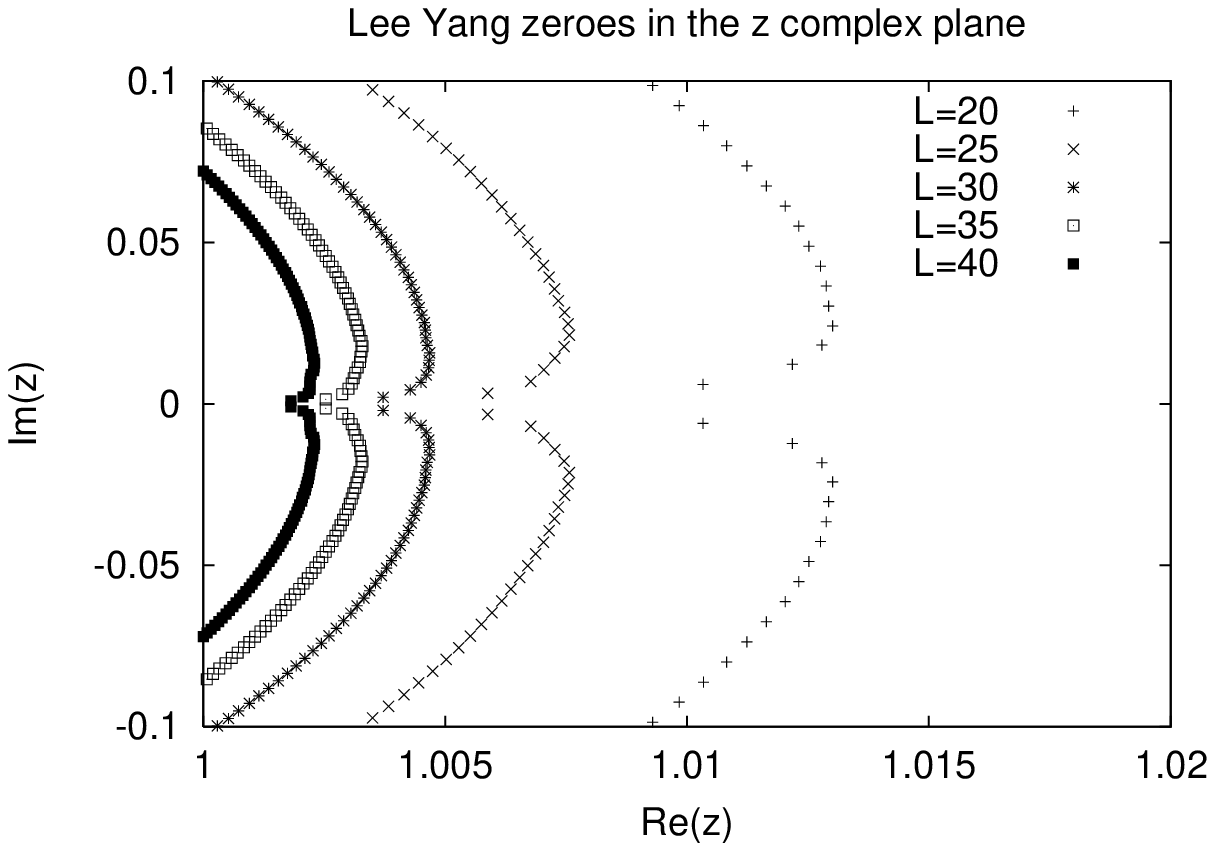}
\end{minipage} 
\vspace{0.3cm}
\caption{Lee Yang zeros in the $z$ plane.
Fig. \ref{patternzeros} a. The curve $\cC_L$ for 
$L=40$. Fig. \ref{patternzeros} b
Local behavior near $z=1$ of the Lee-Yang zeros for various $L$ values
in the $z$ complex plane $Ec=2.2,\epsilon=0.1,h=0$.
\label{patternzeros}}
\ec
\enf

\item For $h=0$ infinitely many  zeros  accumulate on $z=1$ (see Fig. 
\ref{patternzeros} b
and Fig. \ref{dista1}).
Consequently, the pressure is \textit{not analytic
any more in the thermodynamic limit}, as expected.
Note that this property is equivalent to the divergence
of the moments $m_L(q)$ since it can be proved 
that the moments $m_L(q)$, for any $q$ larger
than some $q_0$  diverge if and only if a Lee-Yang
phenomenon occurs. This opens a way toward a scaling theory from
the behavior of Lee-Yang zeros, in a way similar to what was done
in equilibrium statistical mechanics \cite{Janke,Kim,Abe,Zuber,Privman}.

\item For $h > 0$ (sub-critical regime)
the zeros do not pinch $z=1$. This is a clear indication
that in this case there is no critical state in the thermodynamic
limit (Fig. \ref{dista1}). 

\bef
\bc
\begin{minipage}{6cm}
\epsfxsize=6cm
\epsfysize=6cm
\epsffile{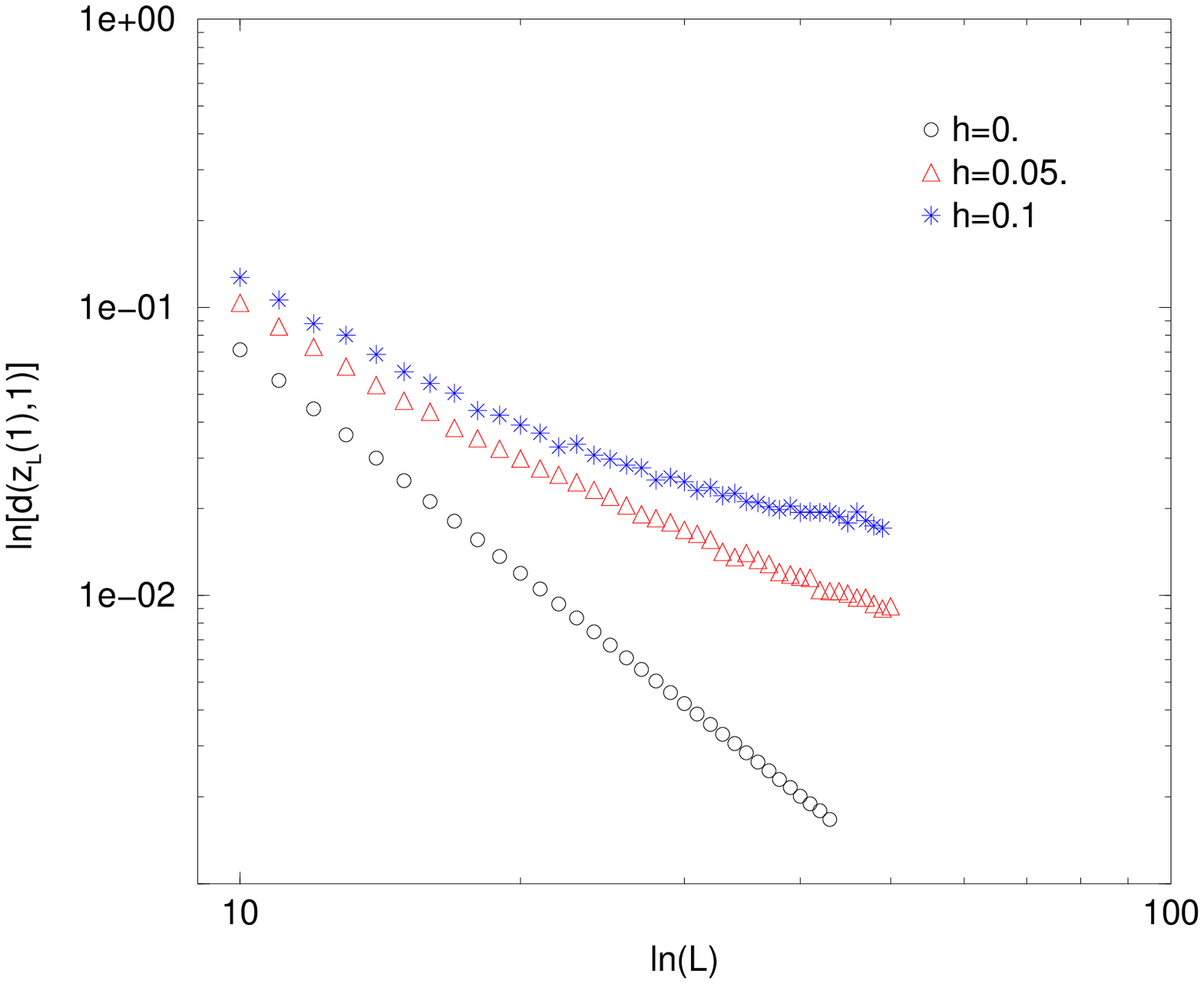}
\end{minipage} 
\vspace{0.3cm}
\caption{Distance of the first Lee-Yang zeros $z_L(1)$ to $z=1$
versus $L$,  $Ec=2.2,\epsilon=0.1$ in the critical case $h=0$
and subcritical case ($h=0.1$, $h=0.05$).
\label{dista1}}
\ec
\enf

\item There  exist  characteristic features of the zeros
that can be connected to the critical behavior. 

\bit

\item For $h=0$, the angle $\theta_L(k)$ (argument of $z_L(k)$)
scales like $\theta_L(k)\sim L^\beta$ (Fig.
\ref{angle}).
It can be proved that, for a probability distribution
converging to a power law and obeying a finite size scaling 
form $s^{-\tau}g(\frac{s}{L^\beta})$, this scaling  
is exact (with however a slight deviation
for the 2 first zeros \cite{CM}).  More generally, 
if the probability distribution obeys the scaling form
(\ref{PLnh}) then one can prove that the angle $\theta_L(k)$
is scales like: 

\beq \label{scalingangle}
\theta_L(k) \sim \frac{2\pi k}{\xi_L^n(h)}
\eeq

For $n=s$, the avalanche size, it follows therefore from (\ref{eL},
\ref{xiL})
that $\theta_L(k) \sim \frac{2\pi k}{\bel^{-\frac{1}{\sigma}}}$
namely that 

\beq \label{fitangle}
\theta_L(k) \sim \frac{2\pi k}{\left[a+cL^{-2}\right]^{-\frac{1}{\sigma}}}
\eeq

\nid where $a,c$ are respectively proportional to
$2(1-\gamma)\epsilon\bpxl$ and $\bpxl$ (see eq. (\ref{eL})).
$a,c$ depend slightly on $L$ (via $\bpxl$) but they converge 
rapidly to a constant as $L\to \infty$
(see fig. \ref{FigeL}a for the $L$ dependence of $\bpxl$).
Therefore, studying the $L$ dependence of $\theta_L(k)$
gives a straightforward way to compute $\sigma$
(resp. $\beta=\frac{2}{\sigma}$). We performed a numerical
study of the angles $\theta_L(k)$ for $E_c=2.2,\epsilon=0.1$
in the conservative and non conservative case ($h=0.1$),
for $L=10-50$. We obtained $\beta=2.59 \pm 0.04$
and $\sigma=0.772$. This corresponds to a critical
exponent $\tau=1.227$, not so far away from the theoretical
value $\tau=1.253$ obtained by the renormalization group analysis
\cite{Pietronero}. Going to larger size would certainly
improve the accuracy, though one has to be careful to pathological
bias induced by the standard numerical procedure where
the number of sample is fixed independently of the system
size \cite{CM}. In Fig. \ref{angle} we plotted the angle $\theta_L(k)$
for $k=4,6,8$ and a non linear fit where the constants
$a,c$ in (\ref{fitangle}) where used as fitting parameters. 
We found $a=0.00439 \pm 2.10^{-3}$ and $c=1.684 \pm 3.10^{-3}$

\bef
\bc
\begin{minipage}{6cm}
\epsfxsize=6cm
\epsfysize=6cm
\epsffile{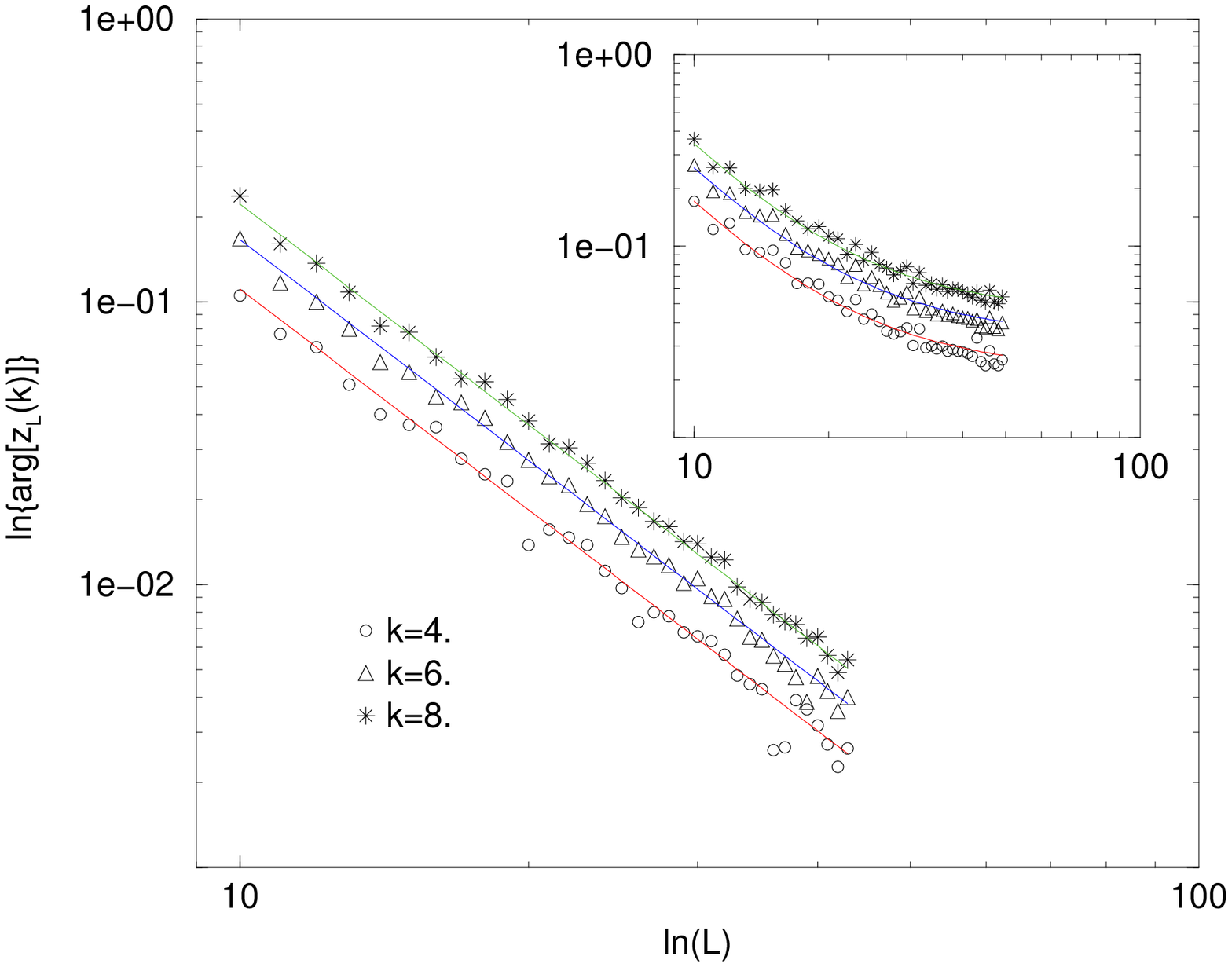}
\end{minipage} 
\vspace{0.3cm}
\caption{Argument of the Lee-Yang zeros $z_L(4),z_L(6),z_L(8)$
for  $Ec=2.2,\epsilon=0.1$ in the critical case $h=0$
and in the subcritical case $h=0.1$.(inset). In
full color lines are drawn the interpolation curves
$y=2\pi k (cx^{-2})^\frac{1}{\sigma}$ (conservative case)
and $y=2\pi k (a+cx^{-2})^\frac{1}{\sigma}$ (non conservative case)
where $a,c$ have been determined by non-linear fitting.
\label{angle}}
\ec
\enf

\item In many models of statistical mechanics the singular part of the 
free energy
obeys a finite size scaling form 
$f^s(t,V)= \frac{1}{V} W[t(AV)^{\frac{1}{2-\alpha}}]$ (equivalent
to (\ref{Kada})) \cite{Fisher,Zuber,Privman},
and the first zeros in
the $t$ plane ($t=log(z)$)  do a characteristic angle with the real
axis, independent of $L$, which can be related to the specific heat
exponent $\alpha$. Here the angle in the $t$ plane \textit{violates the usual scaling
and depends weakly on $L$} (in fact the violation is logarithmic
\cite{CM}). It can be proved that this effect
\textit{is not an indication that  the topological pressure 
does not obey  finite size scaling}
 but is simply due to the value of
the exponent $\tau > 1$.  

\eit

\eit

\ssu{Initial conditions sensitivity.}\label{FDT}

The presence  of a singularity set $\cS$ for the dynamics
induces 
an  interesting phenomenon. If a trajectory approaches
$\cS$ sufficiently close at a time $t$ say,
 a small perturbation in the energy configuration at time $t$
can induce a response \textit{which is not proportional to the perturbation}.
This happens if the perturbed trajectory crosses the singularity
set since the avalanche will be different in the initial
configuration and in the perturbed configuration. 
This effect is obviously due
to the presence of a threshold in the dynamics
definition. More precisely, if in the trajectory of an energy configuration
$\bX$ a site $i$ is such that its energy is arbitrary close to $E_c$
at some time $t$, then obviously a small perturbation on this
site can induce a completely different evolution.
This phenomenon is particularly prominent
if the measure of any $\eta$-neighborhood of  
the singularity set 
is positive, for $\eta >0$. Indeed, in this case,
 with positive probability a trajectory will show non linear
sensitivity to arbitrary weak perturbations
for those times when it approaches the singularity set.
In other words, in this case,
taking two arbitrary close typical energy configurations and exciting
 the same sites along the whole trajectory of these two configurations,
 there will exist a finite time
such that each configuration undergoes a different avalanche.
 This will  result
 in an effective unpredictability
of the evolution (weak initial condition sensitivity) and a failure
of differentiability, or in more physical terms a failure of linear response,
of the observables along typical trajectories. 

It is therefore instructive to compute numerically the probability that a
site is close to $E_c$ within a distance lower than some
$\eta >0$. Call $\cU_{\eta}(\cS) = \left\{
\bX | \exists i, \ |X_i-E_c| < \eta   \right\}$.
Then $\mu_L\left\{\cU_{\eta}(\cS) \right\}$ is exactly the probability
that, during its evolution, a
site is close to $E_c$ within a distance lower than 
$\eta$. In Fig. \ref{muoissing} we have drawn the variation of 
$\mu_L\left\{\cU_{\eta}(\cS) \right\}$
as a function of $E_c$ for $L=30$ (Fig. \ref{muoissing}a) and 
as a function of $L$
(Fig. \ref{muoissing}b), for $\eta=0.001$ and $\eta=0.0001$, $E_c=2.2,\epsilon=0.1,h=0$.
 The statistics were done for $10$ trajectories
and $10^6$ time steps per trajectory.  

\bef
\bc
\begin{minipage}{6cm}
\epsfxsize=6cm
\epsfysize=6cm
\epsffile{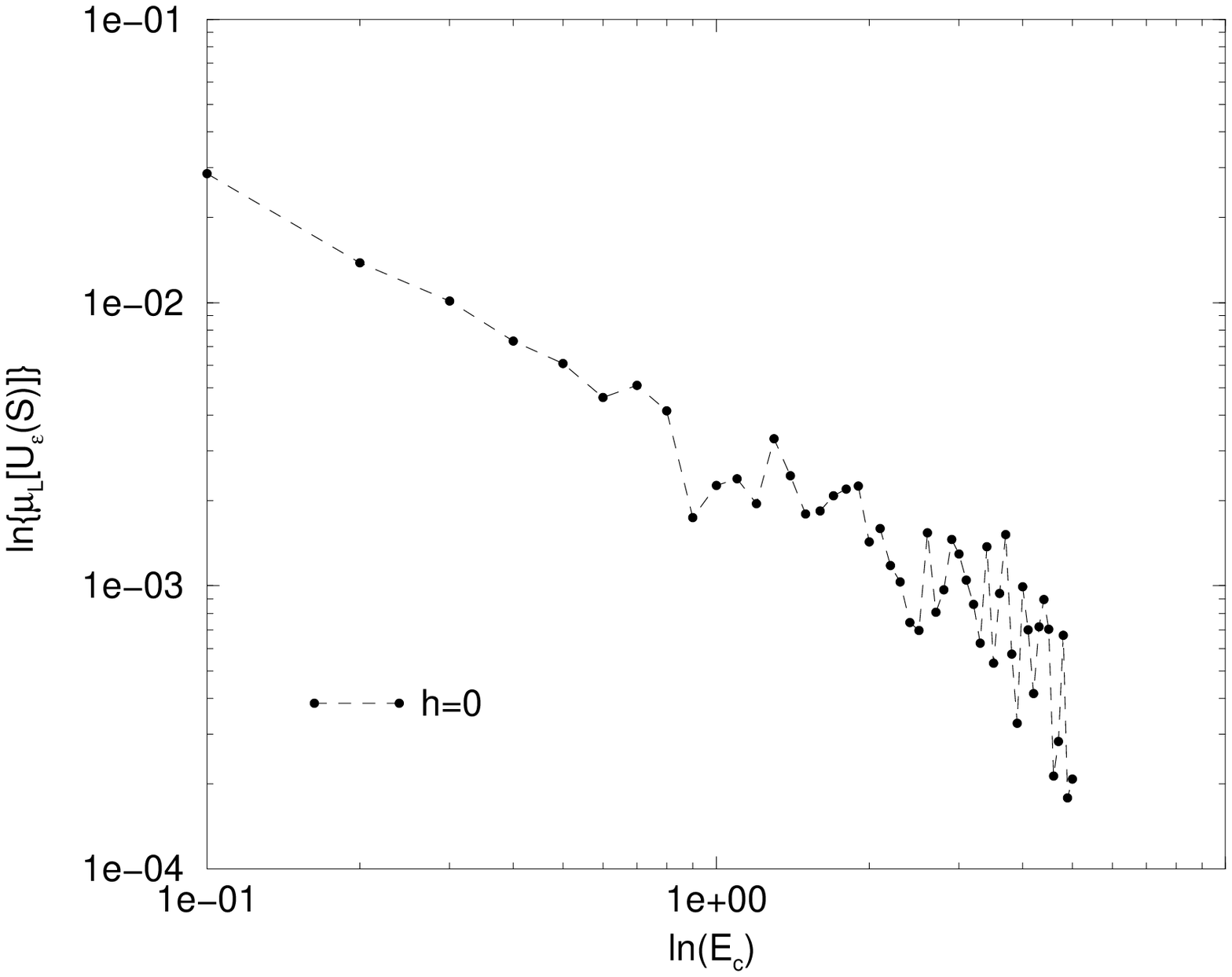}
\end{minipage} 
\hspace{1cm}
\begin{minipage}{6cm}
\epsfxsize=6cm
\epsfysize=6cm
\epsffile{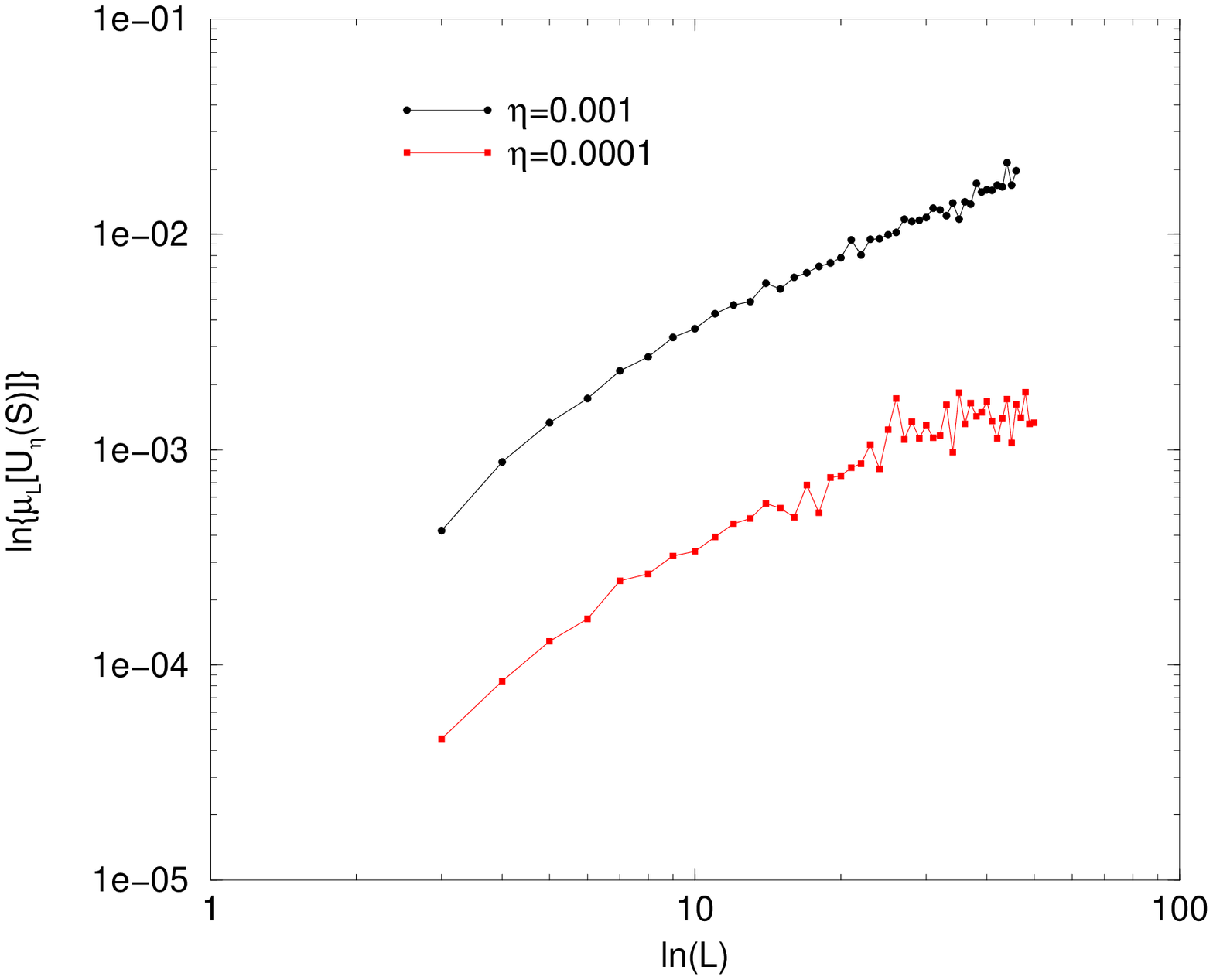}
\end{minipage} 
\vspace{0.3cm}
\caption{Probability $\mu_L\left\{\cU_{\eta}(\cS) \right\}$
that, during its evolution, a
site is close to $E_c$ within a distance lower than 
some $\eta$, as a function of $E_c$,
for $L=30,\epsilon=0.1,h=0$ (Fig. \ref{muoissing}a) and as a
function of $L$,
for $E_c=2.2,\epsilon=0.1,h=0$ (Fig. \ref{muoissing}b).
\label{muoissing}}
\ec
\enf

In  Fig. \ref{muoissingeta} 
we have plotted $\mu_L\left\{\cU_{\eta}(\cS) \right\}$ as a function of 
$\eta$ for $L=30,E_c=2.2,\epsilon=0.1,h=0, h=0.1$. Though we were very careful
in computing this probability (increasing the number of samples when $\eta$
decreases) we believe that the left most points are biased. Ignoring this
point we observed that $\mu_L\left\{\cU_{\eta}(\cS) \right\}$ decay like
$\eta^\alpha$, as $\eta \to 0$, where $\alpha=0.98 \pm 0.01$ for $h=0$
and $h=0.1$. This is supported by another well known
result, already in the early paper of Zhang. The projected energy density
on any site has a characteristic structure depicted Fig. \ref{muoissingeta} b.
More precisely, in this figure we have plotted 
$\rho(E) \deq \mul\left[\bX \ | \ \exists i \in \Lambda, \ X_i \in [E-\iota,E+\iota] \right]$.
The remarkable peaks structure of this distribution has interested many
people but nobody has been able, up to now, to give an analytic form for it.
However, this is  not the point that interests us. Rather, if we focus
on a small neighborhood of $E_c$, we note that the $\rho(E)$ is seemingly
absolutely continuous on the left (explaining the exponent $\alpha \sim 1$).

\bef
\bc
\begin{minipage}{6cm}
\epsfxsize=6cm
\epsfysize=6cm
\epsffile{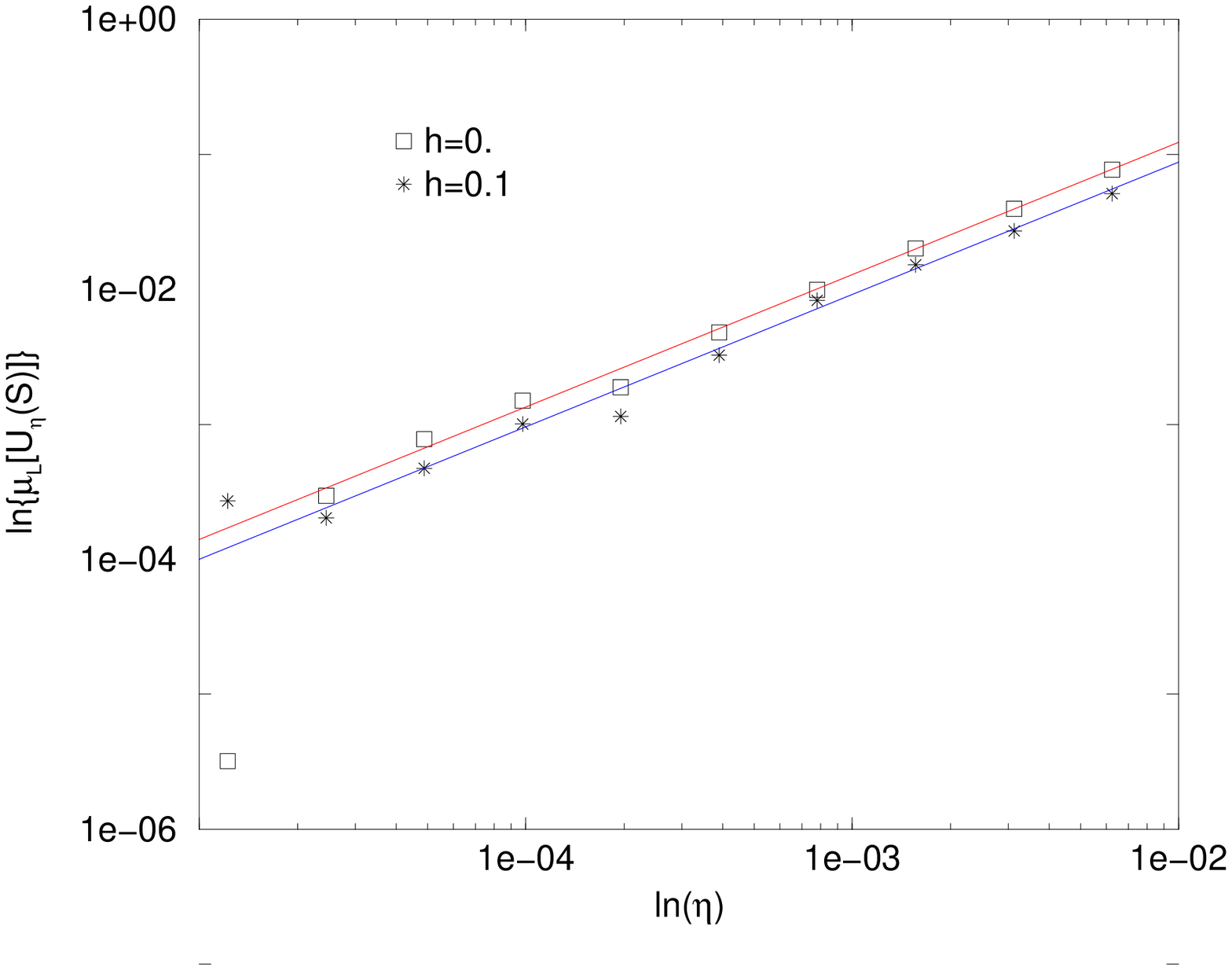}
\end{minipage}
\hspace{1cm} 
\begin{minipage}{6cm}
\epsfxsize=6cm
\epsfysize=6cm
\epsffile{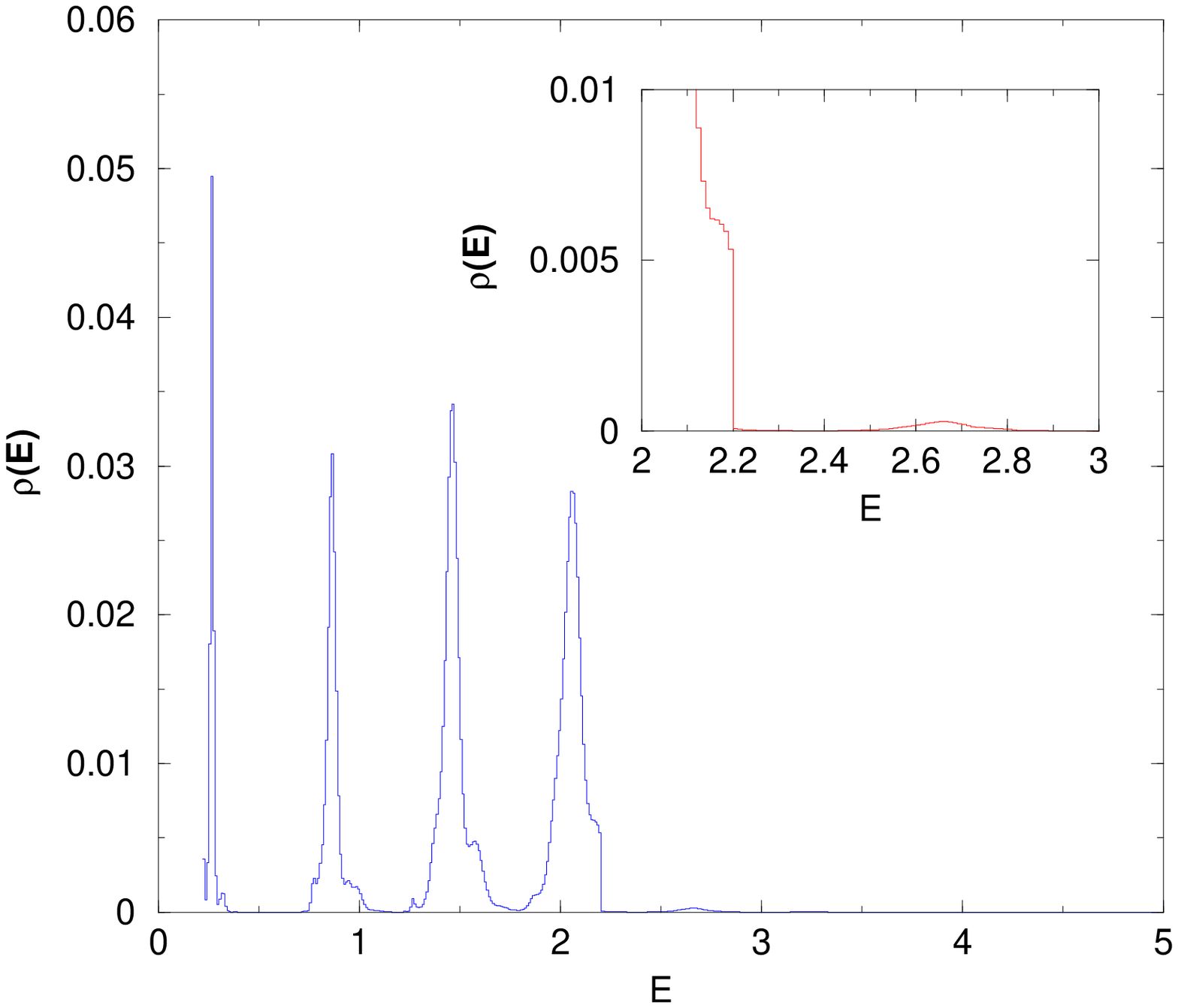}
\end{minipage}
\vspace{0.3cm}
\caption{a. Probability $\mu_L\left\{\cU_{\eta}(\cS) \right\}$
as a function of $\eta$
for $L=30,E_c=2.2,\epsilon=0.1,h=0,h=0.1$.
Fig. \ref{muoissingeta} b. Probability $\rho(E) \deq \mul\left[\bX \ | 
\ \exists i \in \Lambda, \ X_i \in [E-\iota,E+\iota] \right]$,
where $d=2,\ L=30,\  E_c=2.2, \ \epsilon=0.1,h=0, \ 
\iota=10^{-6}$. 
Insert: zoom about
$E_c$.
\label{muoissingeta}}
\ec
\enf

The 
conclusions are quite interesting. On one hand, this last result 
numerically validates
the assumption (\ref{majormu}) we made in section \ref{Statstat} to infer 
the existence of a finite
Markov partition. On the other hand, it shows that the probability to 
approach
a small neighborhood of $\cS$, though weak, is non zero. Consequently,
the Zhang model displays this interesting form of initial condition 
sensitivity
and linear response violation. It would be interesting to adapt this
analysis to other models, in particular those relevant for earthquakes.
Moreover, $\mu_L\left\{\cU_{\eta}(\cS) \right\}$ decreases with $E_c$.
Finally, 
$\mu_L\left\{\cU_{\eta}(\cS) \right\}$
increases with $L$. This suggests that this phenomenon is more and more 
prominent as $L$ grows.
Note, that the number of singularity planes diverges as $L \to \infty$, 
for
$h=0$. It might then  be that the singularities become dense, leading 
to an
extremely impredictible and singular system.

\su{Discussion and conclusion.}

In this paper we have discussed the application of  thermodynamic formalism
to the Zhang's model of Self-Organized Criticality. We have
shown that under physically natural assumptions like ergodicity  Gibbs
measures can be defined to characterize various statistical properties of
the finite size SOC state.
 This opens perspectives to build the equivalent
of the statistical mechanics theory of critical phenomena for SOC models.
It may open the way toward a general setting in which concepts like
universality classes could be properly defined. 
Indeed, though the extrapolation of the thermodynamic formalism to the infinite
lattice size limit needs further developments, this work suggests that the topological
pressure can  be used as an indicator of a phase transition.
In particular we exhibited a Lee-Yang phenomenon for a quantity derived from
the pressure and we noticed that several characteristic patterns emerges,
which could allow a classification of the models in the future.

We would like to discuss now some points that have not been developed
in the paper. 

\ben

\item \textit{Extensions of the thermodynamic formalism.}
In this paper we have focused on the most common 
potentials which are directly related
to  dynamical properties and to the fractal structure
of the support of the invariant measure. We also shown that
they are related to the avalanche size. We now intend to
construct more general potentials allowing on one hand to investigate
the properties of other avalanche observables. On the other hand,
we showed numerically in \cite{BCK4} that the Lyapunov spectrum
exhibits a finite size scaling property with an universal exponent
$\tau_\lambda$ related to the anomalous diffusion exponent.
It would be worth to show this property analytically by 
producing a suitable potential. Finally, we would like to study
the action of the dynamical renormalization group analysis
developed in \cite{Pietronero} on those potentials.

\item \textit{Spectral gap vanishing.} 
In the finite system, the exponential correlation decay 
along the time trajectory is given by the spectral
gap between the largest eigenvalue of the
Markov matrix $\cW$ (which is $1$) and
the second eigenvalue. As mentioned above,
this gap is positive whenever $\cA$
is mixing. 
When $L$ diverges, $\cN_L$ diverges for $h=0$ since
 $\cN_L \geq \xi_L^s$. On the other hand,
the notion of critical phenomena somehow involves
 non exponential correlation decay or a divergence
of the time correlation length which is the inverse
of the spectral gap. Consequently, one
expects that, for
$h=0$ the spectral gap vanishes in the thermodynamic
limit. It might be useful to compute the spectral gap
and particular its $L$ dependence. This could be achieved
from standard techniques on Markov chains \cite{Benaim,Salof},
 provided we have additional informations about the structure of
$\cW$. From the analogy with critical phenomena we expect
the gap to vanish like $L^\eta$ where $\eta$ plays the role
of the exponent giving the spatial correlation decay in statistical
mechanics. This could be a new exponent that might be related to the
exponent $\tau_\lambda$ that we have exhibited in \cite{BCK4}. 

\item \textit{Explicit form of the energy density.} In the section 
\ref{Statstat}
we used the result $\mu_L\left\{\cU_{\eta}(\cS) \right\} \sim \eta^\alpha$
that was only checked numerically. An analytic computation would be 
certainly
be more satisfactory. It could be done if we could achieve the computation
of the energy profile depicted fig. \ref{muoissingeta}b This would also 
allows
to elucidate its particular peaks shape which is still an open problem
in SOC. This could be achieved from the transport equation developed
in the section \ref{Transport}.

\een

These points are under current investigations.\\

{\bf Acknowledgments.}
This work has been partially supported by the Zentrum fuer Interdisciplinaere
Forschung (ZIF) of Bielefeld (Germany), in the frame of the
projet
"The Sciences of Complexity: From Mathematics to Technology to a Sustainable World".
B.C. warmly acknowledge the ZIF for its hospitality. He also thanks the CNRS
for its support.\\

\Appendix 

In this appendix, devoted to non-specialists, we give a brief summary of the thermodynamic
formalism used in section \ref{Gibbs} to construct
Gibbs measures. Useful references are \cite{Bowen,Ruelle,Keller,PP}

Assume that we have a set of symbols
$\omega \in \cI$ and a transition matrix $\cA$.  
Call $\cX$ (resp. $\cX^+$) the set of 
bi-infinite (resp. right infinite) legal sequences.
Write $[\tom]_n$ for a n-cylinder (this is
the subset of $\cX$ where the sequences have
the same $n$ first digits as $\tom$). 
Denote by $\sigma_\cA$ the shift on $\cX$. 
In the sequel we restrict to the set of right infinite sequence
$\cX^+$ as usual in the frame of thermodynamic formalism \cite{Keller,PP}.
There exists a formal analogy between a 
sequence of $\cX^+$ and a (right-infinite) 
one dimensional chain of Potts-like spins taking values in $\cI$.
The transition matrix $\cA$ acts then as a hard-core like potential
in the sense that, if the spin  at the $t$ th place in the chain
has a value $\omega_t$, the next spin (at the place $t+1$) can only take
values in the subset of $\cI$ such that $\cA_{\omega_t,\omega_{t+1}}=1$.
Other transitions are forbidden. 
This formalism allows in particular to study a class of dynamically 
relevant invariant measures
of the dynamical system as formal analogous to Gibbs measure
in a chain of spins. 

Let $F(\cX^+)$ be the space of H\"older continuous functions,
for a  metric 
$d_\theta(x,y)=\theta^N$ on $\cX^+$,  where $N$
is the largest non-negative integer such that $x_i = y_i, i < N$
 and $0<\theta<1$ \cite{PP,Keller}. 
We call a \textit{potential} an element of $F(\cX^+)$.
A potential has in particular the following property:
 
\beq \label{decay}
var_n(\phi) \leq C\theta^n, \ n \geq 0
\eeq

\nid for some $C>0$, some $\theta \in ]0,1[$, where
$var_n(\phi)= \sup \left\{|\phi(x)-\phi(y)|, \ x_i = y_i, i < n \right\}$.
Note that eq. (\ref{decay}) is the equivalent of
the exponential decrease of the interaction with the distance,
insuring the existence of a thermodynamic limit
in statistical mechanics \cite{Meyer,Ruellestat}.
 A \textit{finite range} potential of order r is
such that $\phi(x)=\phi(y)$ if $x_i=y_i$ for $0 \leq i < r$,
namely the values that $\phi$ takes depend only
on the first symbols (resp. the $r$ first spins). Any infinite
range potential in $F_\theta(\cX^+)$ can be uniformly approximated
by a sequence of r-range potentials.

Set
$S_T\phi(\tom)=\sum_{t=1}^T\phi(\sigma_\cA^t\tom)$
where $\phi$ is a potential. Define
the finite \textit{partition function} by

\beq \label{ZT}
Z_T(\phi) = \sum_{\tom \in \chi^+_T} exp{S_T\phi(\tom)}
\eeq

\nid and the \textit{pressure} (resp. the free energy per spin) by :

\beq
\cF(\phi) = \lim_{T \to \infty} \frac{1}{T}log Z_T(\phi)
\eeq

It can be proved that this  
limit  exists 
provided $\phi$ decays sufficiently fast (see eq.(\ref{decay})),
namely there is no ``phase transition'' 
provided the potential belong to $F(\cX^+)$.

A \textit{Gibbs measure} is an invariant measure
where the potential gives
an  exponential weight to the cylinders.
More precisely, $\mu_\phi$ is a Gibbs measure if
 there exists a $A$ such that for
all $T >0$ and $\omega \in \chi_T^+$

\beq
A^{-1} \leq \frac{\mu_\phi(\left[\tom\right]_T)}
{e^{S_T\phi(\tom)-T\cF(\phi)}} \leq A
\eeq
 This essentially
means that $\mu_\phi$ is exponential with a weight given by the sum
of the values that $\phi$ takes on the orbit of $\omega$. Since $Z_T(\phi) \sim
e^{T\cF(\phi)}$ the measure of a spin chain of length $T$
is $\sim \frac{exp({\sum_{t=1}^T\phi(\sigma^t\tom)})}{Z_T(\phi)}$
and the formal analogy with statistical mechanics is straightforward.

In this setting one also associates to each potential
$\phi$ the \textit{Ruelle operator} $\cL_\phi$, which
is a formal extension of the Kramers-Wannier transfer matrix for a spin chain
\cite{Meyer}. An extension of the Perron-Frobenius theorem for
matrices, due to Ruelle, shows that when $\cA$ is irreducible and aperiodic 
$\cL_\phi$ admits a unique maximal positive
eigenvalue which is equal to the pressure $\cF(\phi)$. The corresponding
left eigenvector is the Gibbs measure $\mu_\phi$. 
The spectrum of $\cL_\phi$  provides  informations about the (strong)
mixing properties of $\mu_\phi$. In particular, the spectral gap between
the largest eigenvalue ($\cF(\phi)$) and the remaining part of the spectrum
determines the dominant exponential decay rate of correlations functions
(resp. decay rate to equilibrium). 

$\mu_\phi$ satisfies also a \textit{variational principle}
analogous to the free energy minimization in statistical mechanics.
Namely, call $h(\mu)$ the entropy of the invariant measure
$\mu$ then the  quantity $h(\mu)+\int \phi d\mu$ admits
a unique maximum for $\mu=\mu_\phi$,
 equal to the pressure $\cF(\phi)$. The maximizing measure  
 is naturally called an \textit{equilibrium state}.
Each equilibrium state for a potential $\phi \in F(\cX^+)$
is a
Gibbs state.  
There exists only one maximum (resp. one equilibrium
state related to the observable $\phi$) 
when $\cA$ is mixing.
This situation corresponds to the absence of phase transition
in statistical mechanics. \\

The pressure, beyond the variational principle, shares others characteristic
with the free energy: it is convex, non decreasing, sub-additive ($
\cF(\phi_1+\phi_2) \leq \cF(\phi_1)+ \cF(\phi_2)$) and 
 this is a generating function for the expectations
with respect to $\mu_\phi$. More precisely, the following can be proved
\cite{PP,Keller}.
Let $\phi,\eta \in F_\theta$, and set $\cP(t) \deq
\cF(\phi+t\eta)$, where $t \in \bbbr$,
then :

\beq \label{aimantation}
\cP'(0)=\left.\frac{d}{dt}\cF(\phi+t\eta)\right|_{t=0}=\int \eta d\mu_\phi
\deq E[\eta]_\phi
\eeq 

\nid where $E[]_\phi$ is the expectation  with respect to $\mu_\phi$.
In the same way :

\beq \label{D2}
\sigma^2_\phi(\eta)=\cP''(0)
=\lim_{T \to \infty} \frac{1}{T} \sum_{t=1}^T 
 E\left[\eta(t)\eta(0)\right]_\phi -E\left[\eta\right]_\phi^2 
\eeq 

\nid where 
$E\left[\eta(t)\eta(0)\right]_\phi -E\left[\eta\right]_\phi^2$ 
is the correlation function of the function $\eta$, at
time $t$, the average being performed with respect to $\mu_\phi$. This
can be generalized to correlation functions between different observables
and to higher order \cite{Gaspard}. The coefficient $\sigma_\phi(\eta)$
characterizes  the average fluctuations of $\eta$ along trajectories
weighted by the measure $\mu_\phi$. Provided $\sigma_\phi(\eta)>0$
the central limit theorem holds, namely the fluctuations are Gaussian
(more precisely $ \lim_{T \to \infty}
Prob\left[\frac{\sum_{t=1}^T 
 \eta(t) -TE\left[\eta\right]_\phi}{\sigma_\phi(\eta)\sqrt{T}} < y\right]
=\cN(y)$ where $\cN$ is the characteristic function of the Gaussian distribution).

From the relation (\ref{D2}) one can extract Green-Kubo transport coefficients from
microscopic quantities \cite{Gaspard,Dorfmann}. \\

According to the choice of the potential
one is able to extract different informations about the statistical
properties of the dynamics. The situation is somehow analogous
to statistical mechanics where the choice of the thermodynamic
potential  corresponds to a different choice
of \textit{ensemble}. However one has a priori an infinite
number of choice for the potential (resp. measure) but
only a few of them 
are physically relevant.

\ed
\\

In the conservative case $h=0$,
the numerical simulations report the following behavior.
Fix some avalanche observable $n$ such as  size, duration, area, etc...
After some transient the dynamics reaches a stationary
state such that the probability distribution of $n$,
 $P_L(n)$, where $L$ is the lattice size,
 exhibits  a power law
over a finite range, with a cut-off corresponding to finite size effects. =
As $L$
increases the power law range increases, leading to conjecture that
a critical state is  achieved in the thermodynamic limit,
namely that $P_L(n)$ behaves like $\frac{1}{n^{\tau_n}}$, as $L \to \infty=
$.
$\tau_n$ is called the {\it critical exponent} for the observable $n$.
It seems commonly admitted in the SOC community
that
a classification of the models can be made by the knowledge
of their  critical exponents (``universality classes'').

The full dynamics (excitation + relaxation) of the Zhang model
can be described as a $N+1$ dimensional 
dynamical system, where the additional  dimension
 encodes the excitation dynamics \cite{BCK4}.
It can be proved  that, for finite $L$, the Zhang's model
is weakly hyperbolic, 
namely all Lyapunov exponents
are bounded away from zero. More precisely, there is one
positive Lyapunov exponent $\bom_L log(L^d)$
where $\bom_L$ is the probability, at stationarity,
to excite a site at
a given time (excitation rate). The
remaining $N$ exponents are strictly negative and characterize the relaxat=
ion dynamics.
In particular, the largest negative Lyapunov exponent scales like
 the dissipation rate,
 which is the energy lost per unit  time and per over critical site\cite{B=
CK4}. 
Hyperbolic systems have in particular exponential correlation decay.
However, for $h=0$,
as $L \to \infty$,  the dissipation and excitation rate both go
to zero, algebraically, with a scaling exponent  related
to the critical exponent of  avalanche duration. Consequently,
the Lyapunov exponents go to zero in this limit. This corresponds
to the critical slowing down (algebraic correlation decay and algebraic 
relaxation to equilibrium instead of exponential)
 expected in a critical systems.
 In the non conservative case, when a stationary regime exists,  a
straightforward extension of the proposition $1$ in \cite{BCK4} 
shows that the Lyapunov exponents remain bounded away
from zero as $L \to \infty$. Indeed, the largest negative Lyapunov
exponent is proportional to the dissipation rate which does not vanish
as $L \to \infty$ for $h \neq 0$ (see eq. (\ref{eL})).
In order to insure stationarity this implies that the 
excitation rate and therefore the positive Lyapunov exponent
has also to stay bounded away from zero as $L\to \infty$. 
Consequently, one does not expect a critical slowing down in this case
but rather an  exponential correlation decay.
This is an hint that a critical state can be achieved
only in the conservative case. Further arguments will be given
in this paper. \\

Let $\widehat{x}=\left(
....x_{-n},......,x_{0.,......,}x_{n},........\right) $ be the symbolic
coding of $x$ induced by the Markov-Partition. The forward orbit of $\wide=
hat{%
x}$ encodes the first coordinate in $\widehat{{\cal M}}$ by definition and
the backward orbit of $\widehat{x}$ encodes the point in the configuration
space ${\cal M}$ .This can be made very explicit in terms of the mappings =
$%
T_{\left( i,j\right) }$ associated with the avalanche $\left( i,j\right) $
.Let $\pi =(\pi ^{+},\pi ^{-})$ be the conjugating mapping between $\Sig=
ma $
the shift space and $\widehat{{\cal M}}$ the phase space.Clearly $\pi
^{+}\left( x\right) =\left( i\left( x_{k}\right) \right) _{k\geq o}$ whe=
re $%
i\left( x_{k}\right) $ is the first coordinate in the avalanche $\left(
i,j\right) $ associated with $x_{k}$ .For $\pi ^{-}$ one has the following
expression which is common in the theory of iterated function systems :

\begin{equation}
\pi ^{-}\left( \widehat{x}\right) =\lim\limits_{n\rightarrow \infty
}T_{\left( i\left( x_{-1}\right) ,j\left( x_{-1}\right) \right) }\circ
T_{\left( i\left( x_{-2}\right) ,j\left( x_{-2}\right) \right) }\circ
.....\circ T_{\left( i\left( x_{-n}\right) ,j\left( x_{-n}\right) \right) =
}%
{\cal M}
\end{equation}

Furthermore the measure $\mu _{B}$ projects down to an invariant measure $=
\mu $ on $\widehat{{\cal M}}$ \ via $\mu =\mu _{B}^{\ast }\circ \pi ^{-1=
}$
with marginals $\mu ^{s}=\mu _{B}^{\ast }\circ \pi ^{-}$ and $\mu ^{u}==
\mu
_{B}^{\ast }\circ \pi ^{+}$.If the system is ergodic it follows from the
ergodic theorem for iterated function systems that $\mu _{B}$ almost surel=
y
one has
\begin{equation}
\lim\limits_{n\rightarrow \infty }\frac{1}{n}\sum\limits_{k=1}^{n}\delta
\left( T_{i_{1}}\circ ...\circ T_{i_{k}}x\right) =\mu ^{s}\text{ \ for }=
\forall x\in {\cal M}
\end{equation}

Here $T_{i}$ is the mapping associated with the excitation of site $i$ and=
 $%
\left( i_{1}...i_{k}...\right) $ is any $\mu _{B}$ typical sequence of
excitations .Equivalently one has $\mu _{B}^{\ast }$ a.s. \ for $\widehat{=
x}%
=\left( ....x_{-n},......,x_{0.,......,}x_{n},........\right) \in \Sigma=
 $
\begin{equation}
\lim\limits_{n\rightarrow \infty }\frac{1}{n}\sum\limits_{k=1}^{n}\delta
\left( T_{\left( i\left( x_{1}\right) ,j\left( x_{1}\right) \right) }\circ
T_{\left( i\left( x_{2}\right) ,j\left( x_{2}\right) \right) }\circ
.....\circ T_{\left( i\left( x_{k}\right) ,j\left( x_{k}\right) \right)
}x\right) =\mu ^{s}\text{ for }\forall x\in {\cal M}
\end{equation}

The last two formulas express the fact that $\mu =\mu ^{s}\times \mu ^{u=
}$
is really the physical observable SBR-measure on $\widehat{{\cal M}}$
.Furthermore the canonical invariant measure corresponds in the
thermodynamic setting to the value $\beta =0$.This follows immediately f=
rom
the well known fact that the measure of maximal entropy of a finite
irreducible Markov-chain is obtained by assigning to each outgoing edge of=
 a
vertex $i$ the transition-probability $\frac{1}{d_{out}\left( i\right) }$
\cite{1}.Since in our case the desired measure is the \ unique measure of
maximal entropy one has to put $\beta =0$ .

When a stationnary state exists
a useful quantity has been introduced by Vepsignani et al. \cite{Vespignan=
i}.
The dissipation rate $e_L$  is the average energy lost per unit time and p=
er
critical site.  These authors studied the scaling behaviour of various qua=
ntiti
es
as $e_L \to 0$. We will do a similar analysis by studying the scaling of
$e_L$ with respect to $L$ and $h$. When $h=0$, $e_L \sim L^{-\mu}$ \cite=
{Vespig
nani},
and $\mu=2$
for the Zhang model \cite{BCK4}. For $h \neq 0$ equation (\ref{Edis})
suggests  a scaling form :

\beq\label{eL}
e_L(h)=a(1-\gamma) + bL^{-\mu}
\eeq

\nid where $a$, which is the total average energy transported in the latti=
ce
per unit time, depends a priori on $L,h$.
$b$ depends on $L,h$ but the $L$ dependence is assumed to be negligible
compared to $L^{-\mu}$ when $L$ grows.
As $L\to \infty$, $e_L$ converges to a finite value
$e_\infty(h) = a_\infty(h)(1-\gamma)$,
 for $h\neq 0$.

 Vespignani et al. proposed
in \cite{Vespignani} the following scaling form for the probability distri=
bution
of avalanches observable:

\beq \label{PLnh}
P_L(n,h)=n^{-\tau}f(\frac{n}{n_c(e_L(h))})
\eeq

\nid where $n_c(e_L(h)$ is the correlation length. For the avalanche size
Vespignani et al. proposed a scaling form $s_c(e_L(h))=e_L(h)^{-\frac{1}=
{\sigma}}$.
Consequently, for $h=0$, the correlation length diverge
like $L^\beta$ where $\beta=\frac{\mu}{\sigma}$
 and $P_L(s) = s{-\tau}f(\frac{s}{L^{\beta}})$. This corresponds
to the Kadanoff finite size scaling form (\ref{Kada}).
For $h\neq0$ $s_c(e_L(h))$, converges to a constant
$\left(a(1-\gamma)\right)^{-\frac{1}{\sigma}}$. In the sequel we show how
the correlation length
$s_c(e_L(h))$ and
the corresponding exponent $\sigma$ can be determined from the behaviour o=
f
the Lee-Yang zeros.\

Another potential, closely related to (\ref{det}) is :

 \beq \label{phi2}
\phi(\tom)=q\log(\tml({\omega_1}))+\beta\log(|det(DT_{\omega_1})|)
\eeq

The corresponding 
partition function writes:

\beq \label{ZIFS}
Z_T(\beta,q)=\sum_{\omega \in \chi^+_T} 
\left(\tml({\omega_1})\dots\tml({\omega_T})\right)^q\left(\epsilon^{s(\ome=
ga_1)}
\dots\epsilon^{s(\omega_T)}\right)^\beta=\frac{1}{\cN_L^{qT}} \sum_{\ome=
ga \in \chi^+_T}
\epsilon^{\beta\sum_{t=1}^T s(\omega_t)}
\eeq 

Since  $\epsilon^{s(\omega_t)}$ is the volume contraction at the $t$ time
step of the dynamics,  (\ref{ZIFS}) has some formal analogy with the 
partition function used to construct the multifractal spectrum of
\textit{conformal} probablistic iterated function system verifying the so
called open set conditions. In this setting,
indeed, the multifractal spectrum is given by the Legendre transform
of $\beta(q)$, where $\beta(q)$ is the unique $\beta$ where the topologica=
l
pressure corresponding to  (\ref{ZIFS}) vanishes \cite{Falconer}. Namely:

\beq \label{PIFS}
\lim_{T \to \infty} \frac{1}{T}\log\left(\sum_{\omega \in \chi^+_T}
 \epsilon^{\beta(q)\sum_{t=1}^T s(\omega_t)}\right)=
q\log(\cN_L)
\eeq

Though the Zhang model is not
conformal (see the Lyapunov spectrum)

It can be argued that, provided $E_c$ is sufficiently small,
there exists a finite Markov partition. 
In a nutshell the (non rigorous) argument relies
on the following dimension estimation. Call $\delta_L(i)$ the partial
dimension \footnote{It is almost surely constant from
the ergodic theorem.} in the Oseledec space $E_i(\hX)$\cite{Young}.
 If one can prove that there exists an $E_c$ domain 
such that $\forall i=1 \dots L^d$,
$\delta_L(i) < 1$, then
the support of $\hml$ is a Cantor
set with large gaps. The hyperplanes forming the singularity
set are continuously moving when $E_c$ is varying,
whereas the support of the invariant measure
does not change on open domains of $E_c$ \cite{BCK3}. Therefore,
in this case, the set of $E_c$ values where the singularity set intersects
the support of the invariant measure
is a non generic set, in a topological and metric sense.
The fractal dimensions  $\delta_L(i)$ are given by the Ledrappier-Young fo=
rmula \cite{LedYou}
$\delta_L(i)=\frac{\cH_L(i)}{|\zeta_L(i)|}$ where $\cH_L(i)$ is
the backward entropy in the $i$ th Oseledec direction. $\cH_L(i)$ is
not known in general, but the total entropy $\cH_L=\sum_{i=1}^N \cH_L(=
i) \leq \log(N)$. 
We have shown in section \ref{} that
 $\zeta_L(1)$ becomes arbitrary large
as $E_c \to 0$. Consequently, all negative Lyapunov exponents diverges. Si=
nce 
$\delta_L(i)\leq\frac{\log(N)}{|\zeta_L(i)|}$, there exists an
$E_c$ value, $E_c^\ast(N)$, such that $\forall E_c < E_c^\ast(N)$,
$\delta_L(i) < 1$. Consequently, from the above discussion
the set of $E_c$ values in $[0,E_c^\ast(N)$ where a Markov partition
exists is generic in a topological and metric sense. 
The main
drawback of this argument is however that $E_c^\ast(N) \to 0$
as $N \to \infty$. Obviously, the value of  $E_c^\ast(N)$
has been obtained by a very rough bound on the entropy,
and certainly better estimates can be found. 

Finally, we would like to note that a finite Markov partition do not
exist if the union of the preimages of the singularity hyperplanes
are dense. Though, the argument above suggests that for sufficiently
small $E_c$ values, this is not the case, one can expect the situation
to become worse and worse as $L \to \infty$, since the number of
singularity hyperplanes diverges.\\

--Message-Boundary-4920--